\documentclass[graybox, envcountsame, envcountchap, envcountresetsect]{svmult}
\newcommand*{\ARXIV}{}%
\ifdefined\ARXIV
\usepackage[colorlinks=true,linkcolor=black,citecolor=black,urlcolor=black,bookmarks,breaklinks=true]{hyperref} 
\else
\newcommand{\href}[2]{#2}
\fi

\usepackage{mathtools}
\mathtoolsset{showonlyrefs=true}

\usepackage{mathptmx}
\usepackage{helvet}
\usepackage{courier}
                     
\usepackage{makeidx} 
\usepackage{graphicx}
\usepackage{subfigure}

\usepackage{multicol}
\usepackage[bottom]{footmisc}
\usepackage{amsmath,amssymb,bm}

\usepackage{url}

\usepackage[numbers]{natbib}

\usepackage{url}

\usepackage{amsmath,amssymb,bm}

\DeclareMathOperator{\per}{per} 

\DeclareSymbolFont{extraup}{U}{zavm}{m}{n}
\DeclareMathSymbol{\vardiamond}{\mathalpha}{extraup}{87} 

\makeindex

\usepackage{ourdeclarations}

\begin{document}

\ifdefined\ARXIV
\title*{Cell shape analysis of random tessellations based on Minkowski tensors}
\else
\title{Cell shape analysis of random tessellations based on Minkowski tensors}
\fi
\author{Michael A. Klatt, G\"unter Last, Klaus Mecke, Claudia Redenbach, Fabian M. Schaller, Gerd E. Schr\"oder-Turk}
\authorrunning{Klatt, Last, Mecke, Redenbach, Schaller, Schr\"oder-Turk}
\institute{
  Michael A. Klatt,
  \at Institut f\"ur Stochastik, Karlsruher Institut f\"ur Technologie, 76128 Karlsruhe, Germany\\
  \email{michael.klatt@kit.edu}
  \and G\"unter Last,
  \at Institut f\"ur Stochastik, Karlsruher Institut f\"ur Technologie, 76128 Karlsruhe, Germany\\
  \email{guenter.last@kit.edu}
  \and Klaus Mecke,
  \at Institut f\"ur Theoretische Physik, Universit\"at Erlangen-N\"urnberg, Staudtstr. 7, 91058 Erlangen, Germany\\
  \email{klaus.mecke@physik.uni-erlangen.de}
  \and Claudia Redenbach,
  \at Fachbereich Mathematik, Technische Universit\"at Kaiserslautern,  Postfach 3049, 67653 Kaiserslautern, Germany\\
  \email{redenbach@mathematik.uni-kl.de}
  \and Fabian M. Schaller,
  \at Institut f\"ur Theoretische Physik, Universit\"at Erlangen-N\"urnberg, Staudtstr. 7, 91058 Erlangen, Germany\\
  \email{fabian.schaller@physik.uni-erlangen.de} 
  \and Gerd E. Schr\"oder-Turk,
  \at School of Engineering and Information Technology, Murdoch University, 90 South Street, Murdoch, WA 6150, Australia\\
  \email{g.schroeder-turk@murdoch.edu.au}
}

\maketitle

\abstract{To which degree are shape indices of individual cells of a tessellation characteristic for the stochastic process that generates them?
Within the context of stochastic geometry and the physics of disordered materials, this corresponds to the question of relationships between different stochastic processes and models.
In the context of applied image analysis of structured synthetic and biological materials,
this question is central to the problem of inferring information about the formation process from spatial measurements of the resulting random structure.
This article addresses this question by a theory-based simulation study of cell shape indices derived from tensor-valued intrinsic volumes,
or Minkowski tensors, for a variety of common tessellation models.
We focus on the relationship between two indices:
(1) the dimensionless ratio $\langle V\rangle^2/\langle A\rangle^3$ of empirical average cell volumes to areas, and
(2) the degree of cell elongation quantified by the eigenvalue ratio $\langle \beta_1^{0,2}\rangle$ of the interface Minkowski tensors $W_1^{0,2}$.
Simulation data for these quantities, as well as for distributions thereof and for correlations of cell shape and cell volume,
are presented for Voronoi mosaics of the Poisson point process, determinantal and permanental point processes, Gibbs hard-core processes of spheres, and random sequential absorption processes
as well as for Laguerre tessellations of configurations of polydisperse spheres, STIT-tessellations, and Poisson hyperplane tessellations.
These data are complemented by image data of mechanically stable ellipsoid configurations, area-minimising liquid foam models,
and mechanically stable crystalline sphere configurations.
We find that, not surprisingly, the indices
$\langle V\rangle^2/\langle A\rangle^3$ and $\langle \beta_1^{0,2}\rangle$ are not sufficient to unambiguously identify the generating process
even amongst this limited set of processes.
However, we identify significant differences of these shape indices between many of the tessellation models listed above.
Therefore, given a realization of a tessellation (e.\,g., an experimental image), these shape indices are able to narrow the choice of possible generating processes,
providing a powerful tool which can be further strengthened by considering density-resolved volume-shape correlations.}


\section{Shape descriptors for random cells}
\label{sec_correlation_pp_tess}

In 1966, Mark Kac~\cite{Kac1966} posed the now famous question ``Can one hear the shape of a drum?''
This question referred to the uniqueness of the spectrum of the Helmholtz equation, i.\,e., the eigenmodes---perceptible as acoustic waves to the ear---with respect to different shapes of the Dirichlet boundary conditions.
In general, the answer to Mark Kac's question is ``No'' as examples of distinct drum shapes exist that give the same spectrum of eigenmodes.
Nevertheless, while not providing a unique characterization of the shape of the drum, the eigenmode spectrum contains substantial information about the shape of the drum.
Given a specific eigenmode spectrum, many drum shapes can be excluded as the possible origin of the sound.
While short of being a unique determinant, several aspects and properties of the shape of the drum can be inferred from an observed eigenmode spectrum.
For example, Mark Kac showed how to some degree we can ``hear'' the connectivity of the drum.
In fact, Mark Kac derived a relationship between the eigenmode spectrum of a drum and its shape quantified by Minkowski functionals or intrinsic volumes~\cite{Kac1966}.
This family of integral geometric measures are also at the heart of this article.

\begin{figure}[t]
  \centering
  \subfigure[][]{ \includegraphics[width=0.48\textwidth]{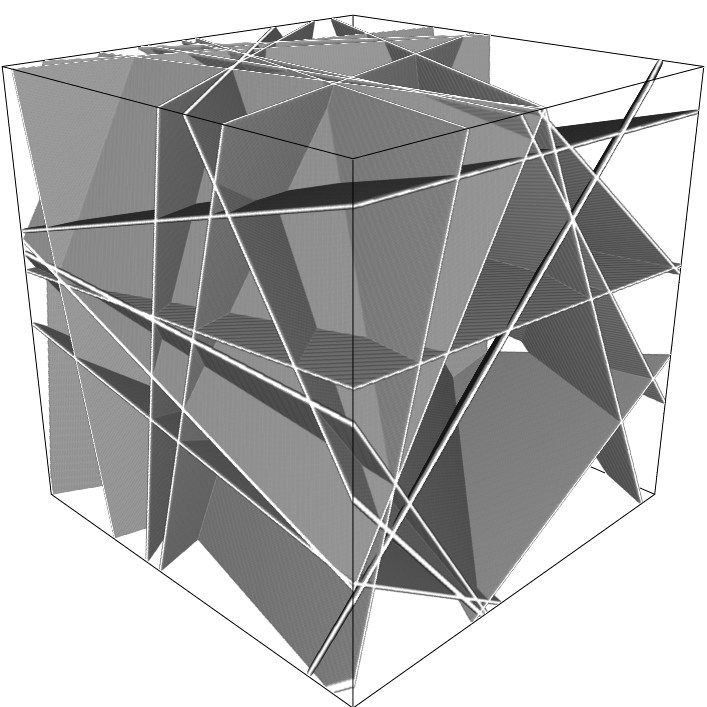} \label{fig:php-not-stit}}
  \hfill
  \subfigure[][]{ \includegraphics[width=0.48\textwidth]{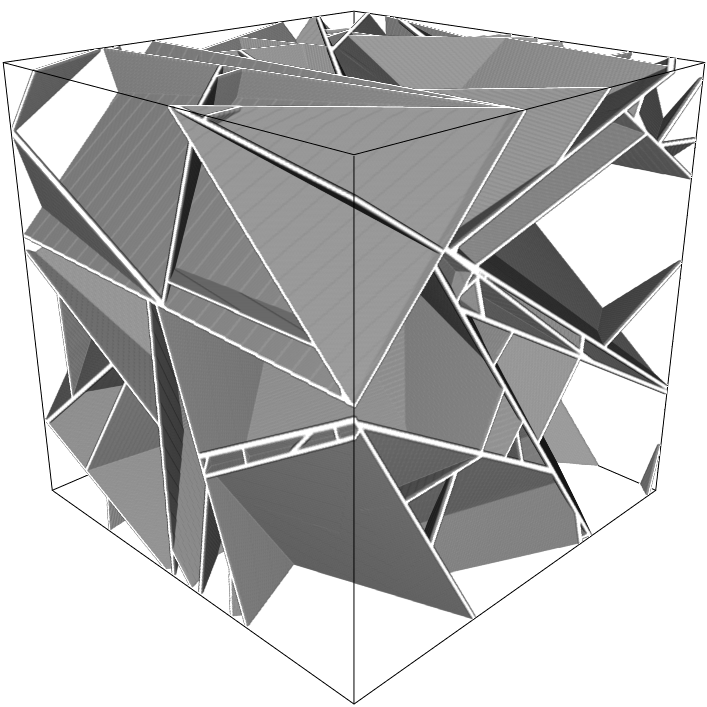} \label{fig:stit-not-php}}
  \caption{Two random tessellations can have identical single cell properties but an obviously different global arrangement, for example,
    a Poisson hyperplane tessellation (a), see Sec.~\ref{sec:Poisshyp}, and a STIT tessellation (b), see Sec.~\ref{sec:STIT-def}.}
  \label{fig:php-vs-stit}
\end{figure}

Here, we address a similar question for stochastic spatial tessellation models, ``Can one 'see' the stochastic process that generates a disordered structure (by considering only geometric characteristics of individual cells)?''.
This question refers to the uniqueness of average or distributional properties of geometric characteristics of a random tessellation --those that can be 'seen'-- with respect to different stochastic processes that underly the formation of the tessellation.
However, the answer to this question bears close analogies to the answer to Kac's question:
the answer to our question is also ``no'', since there are at least two distinct tessellation models that agree in any single cell property.
An example of such a pair is the Poisson hyperplane tessellation~\cite{ChiuEtAl2013} and the STIT tessellation (STable with respect to ITeration)~\cite{NagelWeiss2005}, see Fig.~\ref{fig:php-vs-stit} and Section~\ref{sec:Hyperplane};
a measurement of single cell properties can never uniquely infer which of these two models has generated the tessellation.
On the other hand, like the eigenmode spectrum in Kac's question, the distributions and averages of single-cell properties of the tessellation are strongly dependent on the underlying stochastic process.
In reverse, their measurement can be used to discriminate between possible underlying stochastic processes.

We investigate a variety of random tessellations that are stationary, that is, statistically homogeneous.
We address the question which of their properties can be captured by geometric shape indices of the typical cell.
That is to say, what information is contained in a local structure characterization.

There is a plethora of very different tessellation models that are important across many disciplines, from mathematics, physics, chemistry and biology to computer, life, and social sciences.
Very different types of random or disordered tessellations appear ubiquitously in nature in very different systems, e.\,g., metal alloys, foams, biological tissues, and geological formations.
Moreover, they are not only used to model cellular spatial systems but also for point pattern analysis or local optimization.
For an overview see, e.\,g.,~\cite{Moeller1994, MoellerStoyan2007, OkabeEtAl2009, ChiuEtAl2013}.
In our simulation study, we focus on three-dimensional systems, but the definitions concepts can be applied or generalized to arbitrary dimensions.

Random tessellations can vary both in their geometric construction principle~\cite{Cowan10} and in the underlying stochastic process~\cite{ChiuEtAl2013}.
Although based on the same point pattern, different cells can be constructed according to varying protocols, but the same construction principle can also be applied to different random processes.
We analyze a variety of important mathematical models and physical systems.
We especially compare random tessellations with the same construction principles but different underlying stochastic processes.
How are the shape indices of their typical cells related to each other?
How do they differ from one another?

A special emphasis is on the characterization of anisotropy or elongation of the cell.
Even in a statistically isotropic ensemble, a single cell usually exhibits a non-uniform orientation of its normal vectors.
Similarly as in~\cite{SchroederTurketal2010AdvMater}, we quantify the latter geometric anisotropy 
\ifdefined\ARXIV
by the so-called Minkowski tensors.
\else
by the Minkowski tensors.
\fi
%

An overview of the different construction principles and underlying stochastic processes that are analyzed here is given in Section~\ref{sec:pp_def}.
Here, we also relate the different nomenclatures that are commonly used in various fields of research.
Moreover, we here provide the simulation details.

In Section~\ref{sec:MT}, we present the here used shape indices and discuss how the Minkowski tensors serve as robust and sensitive measures of anisotropy.

In Section~\ref{sec:map}, we estimate the expectations of anisotropy indices of a typical cell in these models and physical systems, from which a ``map of anisotropy'' is constructed that to some extend classifies and relates the different tessellations.
In Section~\ref{sec:distributions}, we extend the analysis and estimate the full probability density functions of both Minkowski functionals and anisotropy parameters.
There, we find that the shape characterization based only on a single descriptor can distinguish such different tessellations as Voronoi or hyperplane tessellations.
However, the probability density functions of the normalized Minkowski functionals for the Voronoi tessellations of (random) point processes can hardly distinguish quite different physical systems.

Therefore, we introduce in Section~\ref{sec:joint-characterization} a more sensitive structure characterization: we determine the mean anisotropy index as a function of the cell volume, i.\,e., conditional on the cell size.
Loosely speaking, we distinguish the shape of small from that of large cells.
For the different physical systems which are indistinguishable w.r.t. the former structure based on a single index, we detect a qualitatively different behavior with the more sensitive analysis based on two different shape descriptors.
The latter can clearly distinguish the different Voronoi tessellations.
Moreover, for the Poisson hyperplane tessellations with an isotropic orientation distribution compared to those with cuboidal cells, we find that while small cells in the latter tessellations tend to be more anisotropic, this trend changes for the large cells that are on average more isotropic.
Section~\ref{sec:conclusions} contains the conclusion.


\section{Construction of random tessellations}
\label{sec:pp_def}

In its most general form, a tessellation is a collection of subsets of $\R^n$ (cells) with pairwise disjoint interiors whose union is $\R^n$.
In such a generality the cells need not be connected, let alone convex.
A rigorous introduction to the mathematical theory of random tessellations with convex cells is given in~\cite{SchneiderWeil2008}; see also~\cite{ChiuEtAl2013}.
Some basic properties of more general (stationary) random tessellations (partitions) are discussed in~\cite{Last06}.

\begin{figure}[t]
  \centering
  \subfigure[][]{\includegraphics[width=0.49\textwidth]{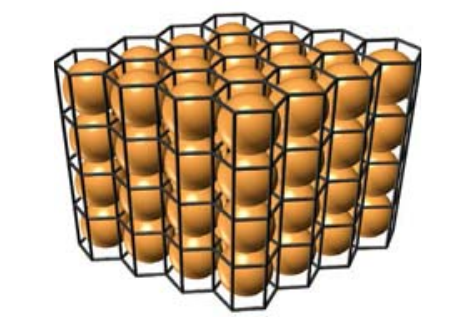}
  \label{fig_vis_AAA}}\hfill
  \subfigure[][]{\includegraphics[width=0.49\textwidth]{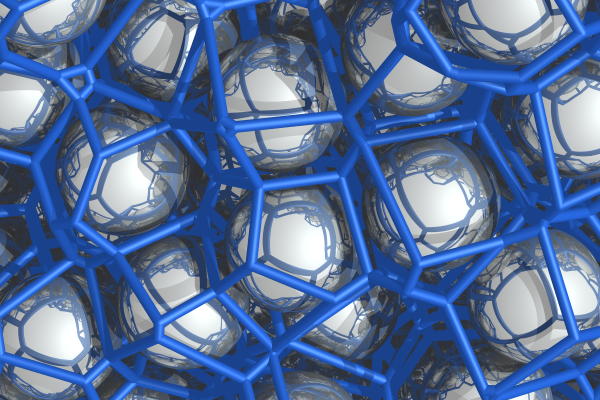}
  \label{fig_vis_equilibrium}}\\
  \subfigure[][]{\includegraphics[width=0.49\textwidth]{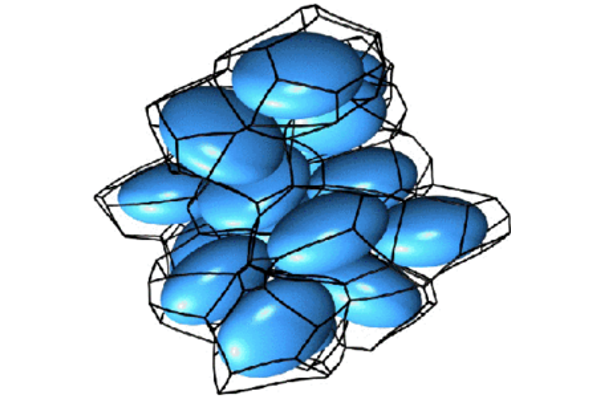}
  \label{fig_vis_ellipsoids}}\hfill
  \subfigure[][]{\includegraphics[width=0.49\textwidth]{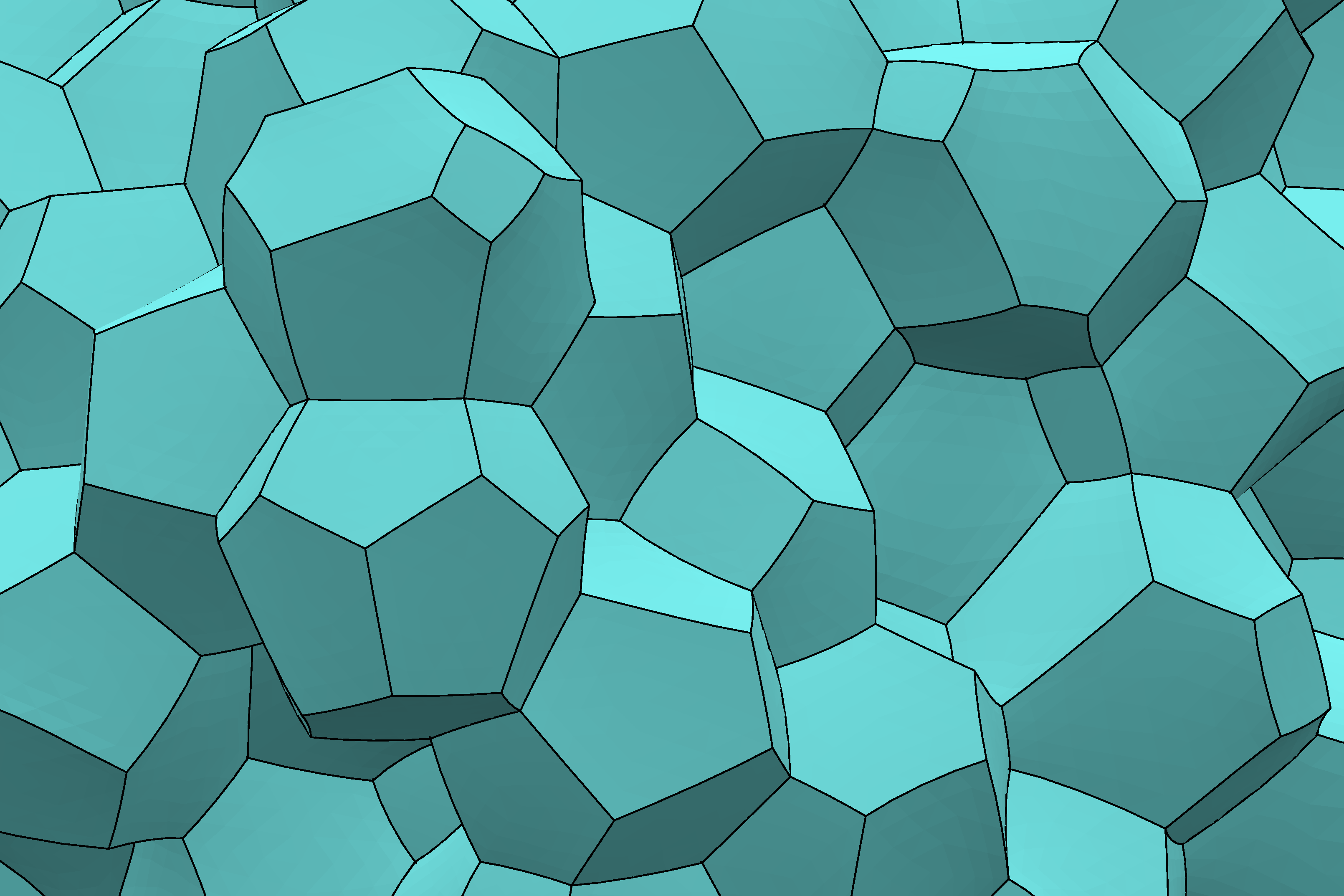}
  \label{fig_vis_foam}}\\
  \subfigure[][]{\includegraphics[width=0.49\textwidth]{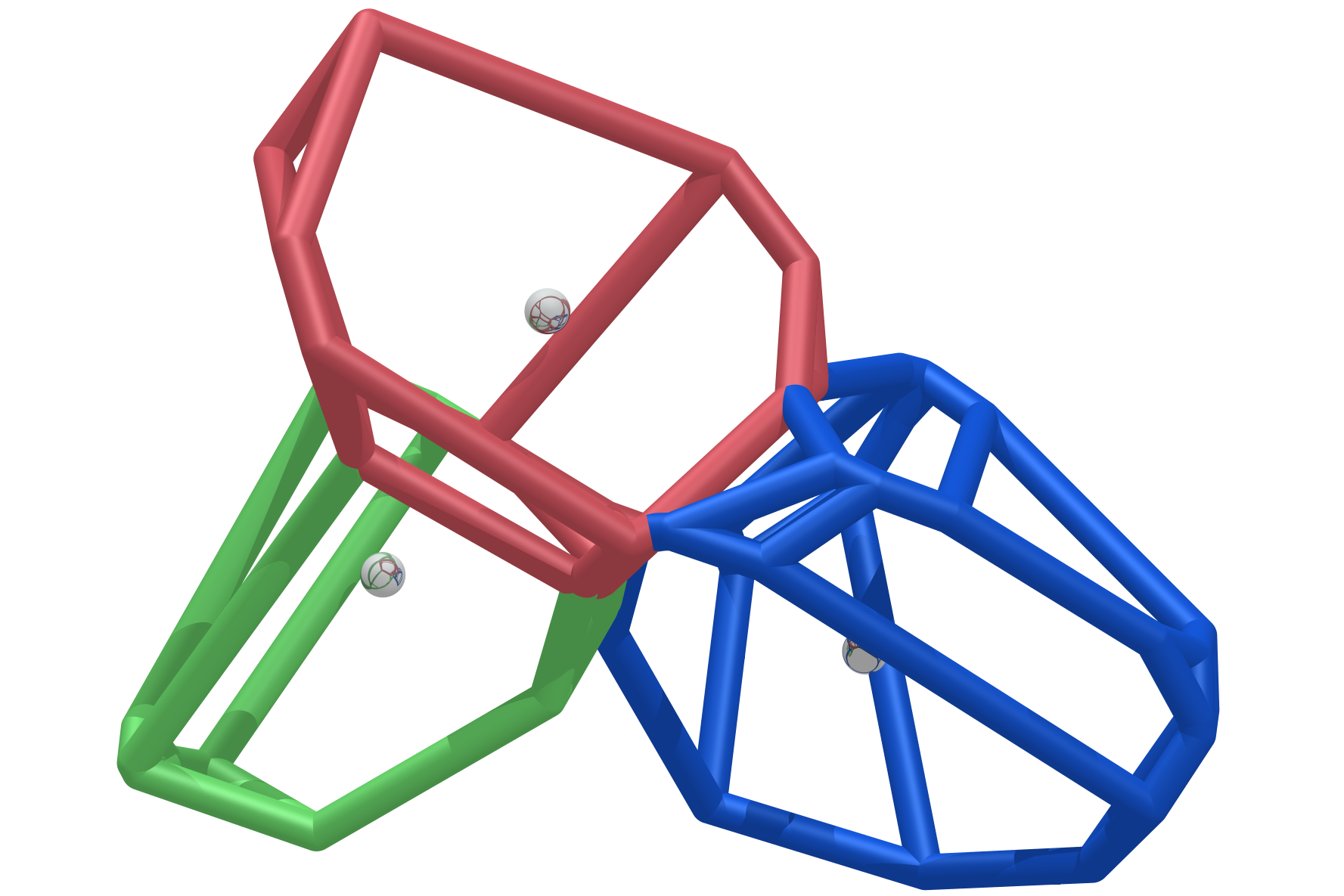}
  \label{fig_vis_Poisson}}\hfill
  \subfigure[][]{\includegraphics[width=0.49\textwidth]{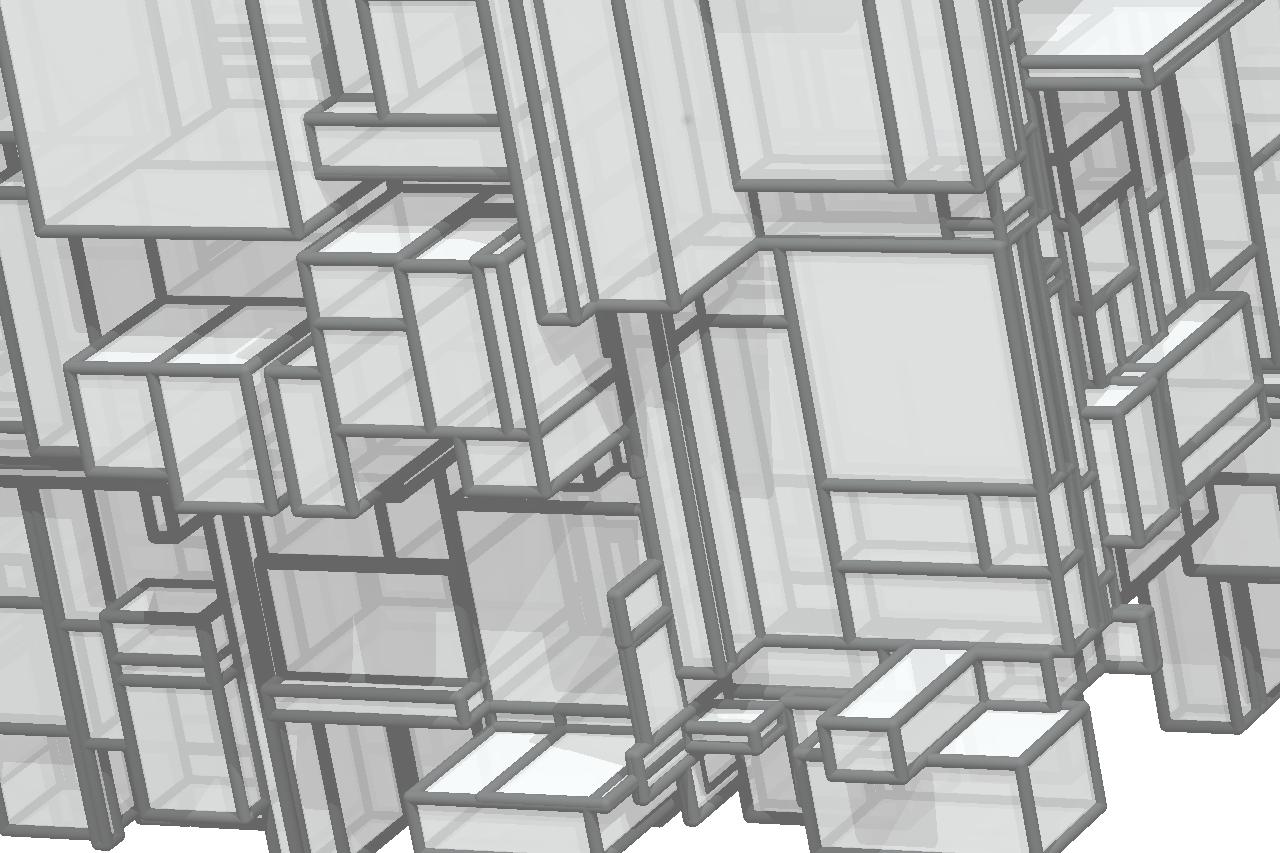}
  \label{fig_vis_STIT}}
  \caption{Examples of tessellations: (a) a crystalline packing of hard spheres
    and the corresponding Voronoi cells of the sphere centers (\href{http://dx.doi.org/10.1063/1.4811939}{reproduced} from~\protect\cite{SchroederTurkSchieleinEtAl2013}, with the permission of AIP Publishing), (b) Voronoi diagram of
    the sphere centers of an equilibrium hard-sphere fluid (using sphere configurations from Steven Atkinson), (c) an experimental random packing
    of ellipsoids and some corresponding set Voronoi cells (\href{http://journals.aps.org/prl/abstract/10.1103/PhysRevLett.114.158001}{reprinted} with permission
    from~\protect\cite{SchallerEtAl2015}. Copyright 2015 by the American Physical
    Society),  (d) random monodisperse foam in the dry limit (data by Andy Kraynik), (e) three
    cells out of a Poisson-Voronoi tessellation, and (f) a STIT
    tessellation.}
    \label{fig_vis}
\end{figure}

The mathematical properties of a random tessellation are the result of a sometimes subtle interplay of the geometric construction principle and the underlying stochastic process.  
Section~\ref{subgeotess} describes some basic geometric construction principles and Sections~\ref{sec:PP}--\ref{subfoam} some stochastic processes driving random tessellations.
Figure~\ref{fig_vis} visualizes some of these examples.

\subsection{Geometric construction principles}
\label{subgeotess}

A Voronoi tessellation (also known as Voronoi diagram) is constructed from a given finite or locally finite subset of points (centers).
The (Voronoi) cell of a center is the set of points in $\R^n$ that are closer to this center than to any other center.
The cells are convex polytopes (finite intersections of half-spaces) which are bounded if the convex hull of the points equals $\R^n$; for examples, see Figs.~\ref{fig_vis_AAA}--\ref{fig_vis_equilibrium}.
Voronoi tessellations can be generalized in several ways~\cite[][Chapter 3]{OkabeEtAl2009}.
For instance one might replace the set of centers by a set of particles of equal or different shape.
This leads to the set Voronoi diagrams.
In such a tessellation, a cell of a particle is the set of points in $\R^n$ that are closer to the surface of this particle than to any other particle.
Instead of the distance to the center of a particle, like in the standard Voronoi tessellation, the distance to the surface of a particle is used.
This definition results in curved facets, see Fig.~\ref{fig_vis_ellipsoids};
for details of the algorithm applied here, see~\cite{SchallerEtAl2013Phil}.
Another rich source for more general models are distances other than the Euclidean one.
A Laguerre tessellation (also called power diagram), for instance, is based on a weighted power distance, where each cell has its individual weight.
Such tessellations do have convex cells.
However, not every center has a non-empty cell.
An example is a system of impenetrable polydisperse spheres, i.\,e., spheres with different volumes, where the radius $R$ of a sphere is the weight for the corresponding cell.
Instead of the Euclidean distance $r$ between a point outside of a sphere and a sphere center, the Laguerre tessellation then uses the weighted power distance $r^2-R^2$.
Voronoi tessellations are widely used in such diverse fields as, e.\,g., astronomy~\cite{feigelson_statistical_1992}, wireless networks~\cite{BaccelliBlaszczyszyn2009a, BaccelliBlaszczyszyn2009b}, archeology, biology, chemistry, computational geometry, geology, marketing; see~\cite{OkabeEtAl2009, ChiuEtAl2013, Torquato2002}.

Tessellations very different from Voronoi diagrams are determined by a (locally finite) collection of (intersecting) hyperplanes.
The interiors of the cells are the connected components of the complement of the union of these hyperplanes.
Hence any such hyperplane is the union of some $(n-1)$-dimensional facets.
The planar case of this hyperplane tessellation is a line tessellation.
The vertices of a non-degenerate tessellation have all degree $2^n$~\cite{SchneiderWeil2008}.

A third class of tessellations results from iterative cell divisions; see~\cite{Cowan10}.
One starts with a (bounded) convex window which is divided by a (random) hyperplane.
Then the cell division process is continued independently on both resulting cells.
In contrast to Voronoi and hyperplane tessellations, this cracking pattern algorithm does not produce face-to-face tessellations, compare Figs.~\ref{fig:stit-not-php} and \ref{fig_vis_STIT}.
In a face-to-face tessellation, the intersection of two cells is either empty or a face of both.
The properties of the tessellation depend on the rules for randomly selecting the cells to be split and the (random) choice of the dividing hyperplanes.
One important example, the STIT-tessellation~\cite{NagelWeiss2005}, is discussed in Section~\ref{sec:STIT-def}, see Fig.~\ref{fig_vis_STIT}.

\subsection{Point processes}\label{sec:PP}

This section considers the completely random Poisson point process, repulsive determinantal point processes, clustering permanental processes,
and patterns with a minimal distance between the points constructed by random sequential addition~\cite{IllianEtAl2007, HKPV06}.

A point process $\Phi$ (on $\R^n$) is a random collection of points which is locally finite, that is, the points are not allowed to accumulate in bounded sets.
The number of points in a (Borel) set $B\subset\R^n$ is denoted by $\Phi(B)$.
Here, point processes are always assumed to be stationary, that is, the distribution of $\Phi$ does not change under simultaneous shifts of all points.
The mean (or expected) number $\gamma=\E[\Phi([0,1]^n)]$ of points falling into the unit cube $[0,1]^n$ is called intensity of $\Phi$.

In most cases the distribution of $\Phi$ can be described by the correlation functions.
This is a countable family of functions $g_k\colon\R^n\to [0,\infty)$, $k\in\N$, satisfying
\begin{align*}
\E[\Phi(B_1)\cdots\Phi(B_k)]=\gamma^k \int_{B_1\times\cdots\times B_k} g_k(x_1,\ldots,x_k)\,d(x_1,\ldots,x_k), 
\end{align*}
for all $k\in\N$ and all pairwise disjoint sets $B_1,\ldots,B_k\subset\R^n$.
Note that the mathematical expectation operator $\E[\cdot]$ corresponds to the ensemble average in physics literature.

For a finite point pattern, $\gamma^k \cdot g_k(x_1,\ldots,x_k)d(x_1,\ldots,x_k)$ can be interpreted as the probability to find a point in each of the infinitesimally small neighborhoods of the positions $x_1,\ldots,x_k$.

\subsubsection{The Poisson process}\label{subPoisson}

Intuitively speaking, a Poisson point process (PPP) is a completely independent point process, where the points are randomly placed in space uniformly distributed, e.\,g., see~\cite{daley_introduction_2003, LastPenrose16}.

The (stationary) Poisson process is the most fundamental example of a point process.
In this case, the correlation functions do not depend on the locations and are simply given by 
$$
g_k(x_1,\ldots,x_k)=1.
$$

For a Poisson process $\Phi$ with intensity $\gamma>0$, the number of points in pairwise disjoint sets are stochastically independent.
Moreover, the number of points in a set $B$ has a Poisson distribution, that is
\begin{align*}
\P(\Phi(B)=m)=\frac{\gamma^m V_n(B)^m}{m!}e^{-\gamma V_n(B)},\quad m=0,1,\ldots,
\end{align*}
where $V_n(B)$ denotes the volume (Lebesgue measure) of $B$.

The PPP can be interpreted as discrete white noise (appearing, e.\,g., as random noise in a detector), and the configurations are equivalent to the grand-canonical version of the ideal gas model.
In the latter case, the intensity is equivalent to the fugacity $z:=\text{e}^{\mu/(k_B T)}$ of the ideal gas (if the unit of length is defined by the thermal de Broglie wavelength),
where $\mu$ is the chemical potential, $T$ the temperature, and $k_B$ the Boltzmann constant.

In this paper we shall be concerned with the Poisson Voronoi tessellation, that is with the Voronoi tessellation generated by the points of a Poisson process, see Fig.~\ref{fig_vis_Poisson}.
Many probabilistic properties of this benchmark model of stochastic geometry are now well understood; see~\cite{Moeller1994, ChiuEtAl2013, SchneiderWeil2008} for first order properties and \cite{LPS14} for (asymptotic) second order properties.

To simulate a Poisson point pattern with intensity $\gamma$, the random number of points within a finite simulation box is drawn from a Poisson distribution with parameter $\gamma\cdot V_{\text{box}}$, where $V_{\text{box}}$ is the volume of the simulation box.
The points are then uniformly distributed within the simulation box.
Data in Figs.~\ref{fig_map}, \ref{fig:epdf-Voronoi}, \ref{fig:epdf-beta-Voronoi}, and \ref{fig:v-beta-Poisson-DPP-RSA} is based on simulations of about 1000 patterns, each with on average 2000 points.

\subsubsection{Determinantal point processes}
\label{sec:dpp}

Determinantal point processes (DPP) were introduced in \cite{Macchi1975} (see also \cite{Berman2014}) to model the behavior of fermions in quantum mechanics.
They are also used to describe transmitters in wireless networks~\cite{DengEtAl2015}.
Determinantal processes model ``soft'' repulsive particles in the sense that although it is unlikely, particles can get arbitrarily close to each other.
(This is in contrast to the hard-sphere systems discussed below).

The mathematical definition of a DPP is based on a kernel $K\colon\R^n\times\R^n\to\R$ which is assumed to generate a self-adjoint, non-negative and locally trace-class integral operator on $L^2(\R^n)$ with eigenvalues in the interval $[0,1]$, see~\cite{shirai_random_2003}.
Then the correlation functions are given by the determinants
\begin{align*}
  g_k(x_1,\ldots,x_k) = \gamma^{-k} \det (K(x_i,x_j))_{1\le i,j\le k}.
\end{align*}
We refer to \cite{HKPV06} for a nice survey of the probabilistic properties of a DPP.
For our present studies we have simulated the DPP with a software package provided by Ege Rubak~\cite{LavancierEtAl2014} written for \textsc{spatstat}~\cite{BaddeleyTurner2005},
which is a package for the statistics software~\textsc{R}.
Data in Figs.~\ref{fig_map}, \ref{fig:epdf-Voronoi}, \ref{fig:epdf-beta-Voronoi}, and \ref{fig:v-beta-Poisson-DPP-RSA} is based on simulations of 100 point patterns with on average 2000 points (using a power exponential spectral model with $\alpha=0.12$ and $\nu=10$, as explained in~\cite{LavancierEtAl2014}).

\subsubsection{Permanental point processes}

Permanental point processes are the attractive counterpart of determinantal processes and can be used to model bosons in quantum mechanics~\cite{Macchi1975}.
The definition of a permanental point process is again based on a kernel $K\colon\R^n\times\R^n\to\R$ which is assumed to generate a self-adjoint, non-negative and locally trace-class integral operator on $L^2(\R^n)$.
Then the correlation functions are given by
\begin{align*}
g_k(x_1,\ldots,x_k)=\gamma^{-k} \per (K(x_i,x_j))_{1\le i,j\le k},
\end{align*}
where the permanent $\per A$ of a $k\times k$-matrix $A=(a_{i,j})$
is given by
\begin{align*}
\per A:=\sum_{\sigma\in\Sigma_k}\prod^k_{i=1}a_{i,\sigma(i)}.
\end{align*}
Here $\Sigma_k$ denotes the group of permutations $\sigma$ of
$\{1,\dots,k\}$.

The mathematical properties of the permanental process are analyzed in \cite{McCullMoe06,HKPV06}, see also~\cite{LastPenrose16}.
A remarkable feature of such a process is that it is doubly stochastic Poisson, i.\,e., an inhomogeneous PPP with a random intensity function.
If the latter is stationary, so is the resulting process.
To explain this we let $(Y_1(x))_{x\in\R^n}$ and $(Y_2(x))_{x\in\R^n}$ denote two independent centered Gaussian random fields with the same covariance function
\begin{align*}
\E[Y_1(x)Y_1(y)]=\E[Y_2(x)Y_2(y)]=\frac{K(x,y)}{2},\quad x,y\in\R^n.
\end{align*}
Let $Z:=(Z(x))_{x\in\R^n}$, where $Z(x):=Y^2_1(x)+Y^2_2(x)$.
Then the permanental process with this kernel $K$ has the same distribution as a point process $\Phi$ with the following two properties.
Given $Z$, the number of points in pairwise disjoint sets are conditional independent, while the number of points in a set $B$ has a conditional Poisson distribution with parameter $\int_B Z(x)\,dx$, that is,
\begin{align*}
\P(\Phi(B)=m\mid Z)=\frac{\left(\int_B Z(x)\,dx\right)^m}{m!}e^{-\int_B Z(x)\,dx},\quad m=0,1,\ldots.
\end{align*}

As pointed out in~\cite{McCullMoe06}, this doubly-stochastic construction allows for a fast simulation of permanental point processes.
First, the two Gaussian random fields $Y_1$ and $Y_2$ are simulated in an observation window $W$.
Then, we simulate an inhomogeneous Poisson point process with intensity function $Z(x):=Y^2_1(x)+Y^2_2(x)$.
Therefore, we simulate a homogeneous PPP $\Phi'$ (see Sec.~\ref{subPoisson}) with intensity $\gamma_0 \geq \max_{x\in W} Z(x)$ but only accept a point $y \in \Phi'$ with probability $Z(y)/\gamma_0$, which is called independent Poisson thinning~\cite{moller_statistical_2003}.

Because of the vast number of possible Gaussian random fields, there is a great wealth of permanental point processes that produce quite different point patterns.
Here, we restrict ourselves to a class of Gaussian random fields that is especially important in physics: the Gaussian random wave model (GRWM), see~\cite{Dennis2007} and references therein.
Intuitively speaking, the GRWM is defined as a superposition of plane waves with random phases and orientations of the wave vector but a constant absolute value of the wave vector;
with an increasing number of random plane waves, the random field converges to a Gaussian random field.
More precisely, for each run we add $N_{w}=100$ plane waves and use
\begin{align}
  f(x) = a_w \cdot \sqrt{\frac{2}{N_{w}}} \cdot \sum_{i=1}^{N_{w}}\cos(k_i\cdot x+\eta_i)\;,
\label{eq_grf_grwm_def}
\end{align}
as a very good approximation of a Gaussian random field.
The random phases $\eta_i$ are uniformly distributed in $[0,2\pi)$.

An anisotropic GRWM can be constructed by choosing a non-isotropic orientation distribution of the wave vectors $k_i$.
Instead of a uniform distribution of the orientation of $k_i$ on the unit sphere, the orientation can, for example, be restricted to a spherical cap with opening angle $\omega$.
Both arbitrarily anisotropic and perfectly isotropic models can be obtained by varying $\omega \in (0,\pi/2]$.

For the present study, we have simulated GRWMs in $W=[0;L)^3$ with $L=25$, $|k_i|=10/L$, $a_w=1/2$, and $\omega=0.01$, 0.03, \dots 0.09, 0.1, 0.3, \dots 0.9, 1.
For each value of $\omega$, we analyze 100 point patterns; each contains on average about 8000 points, but the number of points in the single samples vary between 500 and 40000.

\subsubsection{Random sequential addition} 
\label{sec:rsa}

An intuitive explanation of ``random sequential addition'' (RSA), also called ``random sequential adsorption'' or ``simple sequential inhibition'', is to subsequently place spheres uniformly distributed into space.
However, a sphere is only accepted if there is no overlap with spheres that have already been accepted~\cite{Torquato2002}.
The point process is then formed by the centers of the spheres.
This notion of RSA is nearly identical to the (in mathematics well-known) Mat\'ern III process~\cite{StoyanSchla,MoellerHuberWolpert2010}.
There is only a difference in finite observation windows.
In the latter process, the intensity is fixed and the global packing fraction can (slightly) fluctuate, that is, the fraction of the volume occupied by the particles.
However, in the RSA process studied here the global packing fraction is fixed.
In contrast to the Mat\'ern I and II processes (see e.\,g.\ \cite{Stienen,ChiuEtAl2013,TeichBalBoo13,KidHoe13}) the RSA model does not seem to be amenable to a first and second order analysis.
A likelihood based statistical inference, however, is possible; see \cite{HuWo09}.

Sometimes, RSA is only referring to the saturation limit, i.\,e., a configuration where no additional sphere can be accepted.
The volume fraction in the saturation limit in the spherical spatial case is about $0.384131(\pm 2\cdot 10^{-6})$~\cite{ZhangTorquato2013}.
In this paper, RSA refers to a system for which the global packing fraction is chosen to be some value below this limit. 
Spheres are added to the ensemble only until this global packing fraction is reached.
In the dilute limit, i.\,e., vanishing global packing fraction, an overlap with a previously accepted sphere gets increasingly unlikely.
Thus, the structure of the RSA sphere configurations becomes similar to a Poisson point process.

In chemistry and physics, RSA is used to model the irreversible adsorption or adhesion of, for example, proteins or cells at solid interfaces.
If such a particle is (randomly) attached to the surface, it can no longer move or leave the interface, and it prevents other particles from being absorbed in its neighborhood~\cite{talbot_car_2000, erban_time_2007}.

For Section~\ref{sec:map}, we simulate for each occupied volume fraction ($\phi=0.03$, $0.06$, \dots $0.36$) 100 samples of 2000 spheres.
For Sections~\ref{sec:distributions} and \ref{sec:joint-characterization}, we simulate 1000 patterns where each again consists of 2000 spheres.
A description of the simulation procedure (which follows straightforwardly from the definition of the process) can, e.\,g., be found in \cite[Section 12.3, page 280]{Torquato2002}.

\subsection{Systems of hard particles}
\label{sec:packings-of-hard-particles}

We also analyze systems of hard, impenetrable particles and the resulting Voronoi tessellations: simulations of equilibrium hard spheres, a variety of crystalline arrangements of spheres, and experimentally packed spheres and ellipsoids.

\subsubsection{Equilibrium hard spheres} 
\label{sec:equihs}

If hard spheres are not motionless but allowed to move, the system can equilibrate, so that the point process defined via snapshots of the sphere centers becomes statistically invariant over time, see Fig.~\ref{fig_vis_equilibrium}.
It can then serve as a simple model for liquids.
In mathematics, this model is often called a hard-sphere Gibbs process~\cite{preston_random_1976, daley_introduction_2003, ChiuEtAl2013}.
This is a point process $\Phi$ satisfying the integral
equation
\begin{align*}%
  \label{Gibbs}
\E\bigg[\sum_{x\in\Phi} f(x,\Phi)\bigg] 
=b\cdot\E\bigg[\int f(x,\Phi\cup\{x\})e^{-E(x,\Phi)}\,dx\bigg]
\end{align*}
for all (measurable) functions $f$ of $x$ and $\Phi$.
In physics, it corresponds to a grand canonical ensemble of impenetrable spheres that are in thermodynamic equilibrium with a reservoir that allows for an exchange of energy and particles.
The number $-\log b$ can be interpreted as chemical activity.
The intensity $\gamma$ is an increasing function of $b$, see~\cite[][p.\,189]{ChiuEtAl2013}.
$E(x,\Phi)$ is some fixed positive parameter if the sphere around $x$ does not intersect the spheres around the points of $\Phi$; otherwise $E(x,\Phi):=\infty$.
Here, all spheres have the same radius.
Replacing in the above integral equation $E(x,\Phi)$ by a more general (energy) function, yields Gibbs processes as characterized in \cite{NguyenZessin79}.
The usual definition of a Gibbs process proceeds with specifying the conditional finite window distributions given the configuration in the complement and then using the Dobrushin-Lanford-Ruelle (DLR) equation, see \cite{Ruelle70,ChiuEtAl2013}.

Like for the RSA process, the equilibrium hard-sphere fluid becomes, in the dilute limit, similar to a Poisson point process. It can then be called a hard-sphere gas.
It was shown in \cite{MMSWD01} that the packing fraction of equilibrium hard spheres tends, as $b\to\infty$, monotonically to the closest packing density $\phi_{\mathrm{max}}=\pi/\sqrt{18}\approx 0.74$~\cite{Hales2005}.
With increasing packing fraction, there is a disorder-to-order phase transition~\cite{AlderWainwright1957, WoodJacobson1957} and at a maximum global packing fraction the spheres form a face-centered-cubic crystal (or one of its stacking variants).

This work analyzes equilibrium configurations of hardcore spheres without gravity in the isotropic fluid phase obtained by Monte Carlo simulations.
The data is taken from \cite{KapferEtAl2010}.
The results are averaged over 10 systems per global packing fraction with up to 16384 particles.
Dense systems contain only 4000 particles due to computational costs.

\subsubsection{Crystalline sphere packings}

We also determine the geometric anisotropy of cells in deterministic, perfectly ordered tessellations.
More specifically, we analyze Voronoi cells of crystal lattices that are formed by the centers of hard spheres in mechanically stable packings~\cite{zong_sphere_1999, conway_sphere_1999}.
We compare the random packings of hard spheres to crystalline packings of hard spheres that are isostatic, which means no sphere can be moved while all other spheres are kept fix.
(In three dimensions, each sphere touches at least four other spheres of which not all are on the same hemisphere.)
Moreover, each sphere is connected to any other sphere via a chain of contacts.
The data includes a great range of Bravais lattices, which in three-dimensions are lattices that are generated by discrete translations of three independent vectors, therefore all lattice sites are equivalent.
The most isotropic unit cells (w.r.t. the distribution of the normal vectors on the cell boundary) are in our analysis those of the simple cubic, body-centered, and face-centered cubic packing as well as the hexagonal close-packed arrangement and the hexagonal AAA stacking, see Fig.~\ref{fig_vis_AAA}.

The data for the Minkowski functionals and tensors of the Voronoi cells for these sphere packings are provided by Richard Schielein, for more details see~\cite{SchroederTurkSchieleinEtAl2013}.

\subsubsection{Jammed ellipsoids and spheres}

Jammed packings of hard particles need not to be perfectly ordered like the crystalline sphere packings.
There are also mechanically stable packings that are disordered.
The local structure is more similar to an equilibrium liquid like in Sec.~\ref{sec:equihs} but the particles are jammed.

Moreover, the simple model of hard spheres can be extended to non-spherical particles, namely ellipsoids.
However, in contrast to the equilibrium hard-sphere fluid, we here analyze experiments with non-equilibrium jammed packings.
This work uses oblate ellipsoids (e.\,g. two equally long and one small half-axis) created by 3D printing with aspect ratios $\alpha = 0.40, 0.60, 0.80, 1.00$,
where the aspect ratio is defined as the ratio of the smallest to the largest length of a semi-axis.
For each aspect ratio at least ten packings of 5000 ellipsoids with different global packing fraction are analyzed.
The particles are randomly packed into a cylindrical container by different preparation protocols to get an initial loose packing.
They can then be compactified by tapping to get a variety of different packing fractions~\cite{SchallerEtAl2015}.
A 3D image is gained by imaging the packing with X-ray tomography.
The positions and orientations of the ellipsoids are detected from the grayscale image of the are detected from the grayscale image of the tomographic reconstruction, for more details see \cite{SchallerEtAl2013AIP}.
To reduce boundary effects, the outer particles are removed for the analysis leaving approximately 800 particles in the core region.

The Voronoi diagram of the ellipsoid centers is not a useful tessellation of the void space for non-spherical particles like ellipsoids.
The particles are anisotropic, and facets of the Voronoi cell could cut the particles.
Therefore, we construct the set Voronoi diagram of the jammed ellipsoid packings, see Fig.~\ref{fig_vis_ellipsoids} and~\cite{SchallerEtAl2013Phil}, as described in Sec.~\ref{subgeotess}.

\subsection{Tessellations constructed by hyperplanes}
\label{sec:Hyperplane}

Sections~\ref{sec:PP} and \ref{sec:packings-of-hard-particles} discuss particle processes and their Voronoi tessellations or generalizations thereof.
Quite different tessellations can be constructed by a collection of intersecting hyperplanes.
In Sec.~\ref{sec_correlation_pp_tess}, Figure~\ref{fig:php-vs-stit} compares realizations of two such tessellations,
which have the same distribution of the typical cell, i.\,e., the same single cell properties, but an obviously different global structure.

\subsubsection{Poisson hyperplane tessellations}\label{sec:Poisshyp}

Loosely speaking, a Poisson process $\Phi$ of hyperplanes can be defined by replacing the points of a PPP by hyperplanes~\cite{SchneiderWeil2008}.
This means that $\Phi$ is a countable collection of random hyperplanes with the following two properties.
First, the random number of hyperplanes with a prescribed property follows a Poisson distribution.
Second, given a finite number of mutually exclusive properties, the random numbers of hyperplanes with these properties are stochastically independent.
Such a process is stationary if its distribution does not change under simultaneous translations of the hyperplanes.
In this case, the distribution of $\Phi$ is determined by an intensity parameter
(the cumulative surface area of hyperplanes in the unit volume) and a directional orientation distribution $\Q$, an (even) probability measure on the unit sphere.
If  $\Q$ is uniform, then the tessellation is statistically isotropic.
In general $\Q$ can be used to model preferred directions.

Here, we analyze Poisson hyperplane tessellations (PHP) with two different orientation distributions of the hyperplanes.
They are either isotropically distributed, see Fig.~\ref{fig:php-not-stit}, or only three directions are allowed that are orthogonal to each other, and the probability for each is $1/3$.
Thus, the single cells form cuboids. 

We simulate these tessellations with unit intensity.
For the isotropic tessellations, we use software written by Felix Ballani.
We analyze $2 \cdot 10^7$ cells in the isotropic tessellation
and $10^7$ cuboidal cells in the tessellation with three allowed directions.

\subsubsection{STIT tessellations}\label{sec:STIT-def}

STIT (STable with respect to ITeration) tessellations \cite{NagelWeiss2005,ChiuEtAl2013} are the result of an iterative cell division process.
The original cell  (e.\,g., the simulation box) has an exponential random lifetime whose parameter (in the isotropic case) is proportional to its first intrinsic volume, i.\,e., the $(n-1)$-st Minkowski functional.
At the end of a lifetime, the cell is divided by a random hyperplane.
The two new cells again have exponential lifetimes whose parameters are determined as above and which are assumed to be independent.
Stopping this process after some fixed deterministic positive time yields the STIT tessellation.
Different STIT models can be constructed from different orientation distributions of the random hyperplanes, compare Figs.~\ref{fig:stit-not-php} and \ref{fig_vis_STIT}.
Potential applications of the STIT tessellations are approximations of
'hierarchical' crack or fracture structures~\cite{NagelWeiss2005}. 

Although being globally very different from Poisson hyperplane tessellations,
the typical cell of a STIT tessellation has the same distribution as that of the corresponding Poisson hyperplane tessellations; see~\cite{NagelWeiss2005}.
Therefore, also the joint distribution of all Minkowski functionals and tensors of a typical cell is the same in a STIT or in a Poisson hyperplane tessellation (with the same intensity and orientation distribution of the hyperplane directions).
For further properties of STIT-tessellations we refer to~\cite{MeNaWei11,SchrThaele11,SchrThaele12}.

\subsection{Random dry foam and Laguerre tessellations}\label{subfoam}

Foams like dry soap froth in the limit of a vanishing liquid content~\cite{kraynik_foam_2003}, that is with infinitely thin soap films, are important examples of tessellations which minimize surface area~\cite{hilgenfeldt_accurate_2001}. 
The foam that is analyzed here, see Fig.~\ref{fig_vis_foam}, is a realistic model for monodisperse dry soap froth, where all cells have the same volume~\cite{KraynikEtAl2003}.

Interestingly, for random foams, the basic stochastic and geometric building blocks are closely connected~\cite{Kraynik2006}.
In this sense, it can be seen as a hybrid model of different construction principles and stochastic processes.

The simulation starts from the Voronoi diagram of a random hard-sphere packing, derived by the Lubachevsky-Stillinger packing algorithm~\cite{LubachevskyStillinger1990}.
Then, the soap froth is equilibrated by Kenneth Brakke's \textsc{Surface Evolver}~\cite{Brakke1992}.
The surface area is minimized, and the mechanical forces balanced~\cite{KraynikEtAl2003, KraynikEtAl2004, EvansEtAl2013}.
The data is provided by Andy Kraynik.

We also analyze random Laguerre tessellations with varying volume distributions, which are intended as a mathematical model similar to polydisperse foam structures, where the volume of the cells can vary strongly~\cite{Lautensack2007}.

The data is taken from~\cite{Redenbach2009}.
First, a random ensemble of hard spheres is simulated where the volume of the spheres follows a log-normal distribution with a coefficient of variation (CV) between $0.2$ and $2.0$.
The packing fraction, i.\,e., the volume fraction that is occupied by the spheres, is $60\%$.
For each chosen parameter, we simulate five samples of sphere packings in the unit cube (with periodic boundary conditions) each containing $10000$ spheres.
Using the radii of the hard spheres as weights, the Laguerre tessellation is constructed as described in Sec.~\ref{subgeotess}.


\section{Minkowski tensors and anisotropy indices}
\label{sec:MT}

\ifdefined\ARXIV
Intrinsic volumes are fundamental characteristics of convex and more general sets.
In physical literature, they are commonly referred to as ``Minkowski functionals''
(The only difference to the mathematical literature is in the normalization.)
These functionals and their tensor valuations extensions, the
``Minkowski tensors'' are efficient numerical tools, which have
\else
Intrinsic volumes are in physical literature commonly referred to as ``Minkowski functionals'' and the tensor valuations as ``Minkowski tensors''.
(The only difference to the mathematical literature is in the normalization.)
They are efficient numerical tools, which have
\fi
been successfully applied to a variety of biological~\cite{BeisbartEtAl:2006, BarbosaEtAl:2014} and physical
systems~\cite{Mecke1998, MeckeStoyan2000, Klatt2016} on all length scales from nuclear physics~\cite{SchuetrumpfKlattEtAl2013,
SchuetrumpfKlattEtAl2014}, over condensed and soft matter~\cite{HansenGoosMecke2009, KapferEtAl2012, WittmannEtAl2014, ScholzWirnerEtAl2015}, to
astronomy and cosmology~\cite{KerscherMecke2001, Colombi2000, Schmalzing1999, Gay2012, Ducout2013, GoeringKlattEtAl2013} as well as to pattern
analysis~\cite{Mecke1996, Becker:2003, MantzJacobsMecke2008, ScholzEtAl2015}.
They allow for a versatile morphometric analysis of random spatial structures on very different length scales~\cite{Klatt2016}.

The Minkowski tensors allow for a comprehensive~\cite{Alesker:1999a, Alesker:1999b} and systematic approach to quantify various aspects of structural anisotropy~\cite{SchroederTurketal2010AdvMater}.
A local analysis based on the anisotropy of a single cell quantified by the Minkowski tensors, e.\,g., allows to detect phase transitions and the onset of crystallinity in jammed packings~\cite{KapferEtAl2010, Kapfer2011, KapferEtAl2012, MickelEtAl2013}.

Free software to calculate Minkowski functionals and tensors of both voxelized and triangulated data \textsc{papaya} and \textsc{karambola} (for two and three dimensions, respectively) is available at:\\
\hbox{}\hfill\url{www.theorie1.physik.fau.de/research/software.html}\hfill\hbox{}

A comprehensive introduction to the Minkowski tensors as anisotropy indices and exemplary applications can be found in~\cite{SchroederTurketal:2010jom, SchroederTurketal2010AdvMater}.
A comparison of physical and mathematical notation is provided in~\cite{SchroederTurkEtAl2013NewJPhys}.

\subsection{Integral geometric definition}

\ifdefined\ARXIV
In $n$-dimensional Euclidean space, the Minkowski functionals can be defined by the Steiner formula~\cite{SchneiderWeil2008}.
Given a convex body $K$, i.\,e., a compact, convex subset of $\mathbb{R}^n$, the parallel body $K_{\epsilon}$ at a distance $\epsilon \geq 0$ is the set of all points $x$ for which there is a point $y$ in $K$ such that the distance between $x$ and $y$ is smaller or equal to $\epsilon$.
The volume $W_0$ of this parallel body $K_{\epsilon}$ can be expressed by a polynomial of $\epsilon$
\begin{align*}
  W_0(K_{\epsilon}) = W_0(K) +
  \frac{1}{n}\sum_{\nu=1}^n\epsilon^{\nu}\cdot\binom{n}{\nu}\cdot W_{\nu}(K)\;,
\end{align*}
where $W_0(K)=V_n(K)$ and the coefficients $W_1(K),\dots,W_n(K)$ depend on $K$ but not on $\epsilon$.
The so-called Minkowski functionals $W_1(\cdot),\dots,W_n(\cdot)$ can be generalized to finite unions of convex bodies, so that
\begin{align}
  W_{\nu}(K\cup L) =  W_{\nu}(K) + W_{\nu}(L) - W_{\nu}(K\cap L)\;. 
\end{align}
This is the additivity property~\cite{SchneiderWeil2008}.
By our choice of normalization, the zeroth order Minkowski functional of an $n$-dimensional unit ball $B^n=\{x\in\mathbb{R}^n\mid\|x\|\leq 1\}$ is equal to its volume $W_0(B^n)=:\kappa_n$, and the higher-order Minkowski functionals are equal to its surface area $W_{\nu}(B^n)=n\kappa_n$ ($n\geq \nu\geq 1$).
As mentioned above, in the mathematical literature the intrinsic volumes $V_{\nu}$ are more commonly used than the Minkowski functionals, but only differ by a proportionality constant and the order of the indices
\begin{align*}
  V_{\nu}(K) &:=
  \frac{1}{n\cdot\kappa_{n-\nu}}\cdot\binom{n}{\nu}\cdot
  W_{n-\nu}(K)\quad\mathrm{for\ }\nu\leq n-1\;.
\end{align*}

The Minkowski tensors of a convex body $K$ in the $d$-dimensional Euclidean space can be defined using the so-called support measures $\Lambda_{\nu}(K;\cdot)$~\cite{SchneiderWeil2008}, which can in turn be defined by a local Steiner formula.
Their total mass yields the intrinsic volumes, that is $V_{\nu}=\Lambda_{\nu}(K;\mathbb{R}^n\times\mathbb{S}^{n-1})$.
Like the scalar functionals, the tensors can be generalized to non-convex bodies by using their additivity~\cite{SchneiderWeil2008}.
For consistency with the Minkowski functionals, we use for $\nu\geq 1$ a different normalization $\Omega_{\nu}(K;\cdot):=\frac{n\cdot\kappa_{n-\nu}}{\binom{n}{\nu}}\Lambda_{n-\nu}(K;\cdot)$.
The Minkowski tensors are then defined as~\cite{McMullen1997, SchneiderWeil2008}
\begin{align*}
  W_{0}^{r,0}(K)&:= \int_{K}\vec{x}^r\text{d}^n\,\vec{x} \;,\\
  W_{\nu}^{r,s}(K)&:= \int_{\mathbb{R}^n\times\mathbb{S}^{n-1}}\vec{x}^r\vec{u}^s\,\Omega_{\nu}(K;\text{n}(\vec{x},\vec{u})) \quad\mathrm{for\ }n\geq \nu \geq 1 \;,
\end{align*}
where $\vec{x}^r$ or $\vec{u}^s$ are symmetric tensor
products, and $\vec{x}^r\vec{u}^s$ is the symmetric tensor product of the
tensors $\vec{x}^r$ and $\vec{u}^s$.
\else
To allow easy comparison of the results presented here with the existing physical literature,
we shortly recall the definitions of these quantities with the relevant notational adaptations;
see Sec.~\ref{01-sec3} in Chapter \ref{chap-01} for details.

Let $K$ be a convex body in $\R^n$, that is, a compact convex subset of $\R^n$.
The parallel body $K+\rho B^n$, $\rho\ge 0$, consists of all points in $\R^n$ with Euclidean distance at most $\rho$ from $K$.
The volume $V_n(K+\rho B^n)$ of the parallel body $K+\rho B^n$ can be expressed by the following version of Steiner's formula:
\[
  V_n(K+\rho B^n)=W_0(K)+\frac{1}{n}\sum_{\nu=1}^{n}\rho^\nu \binom{n}{\nu} W_{\nu}(K).
\]
A comparison with the intrinsic volumes in Eq.~\eqref{01-3.4} shows that $V_n=W_0$ is the volume and that
\[
  V_\nu=\frac{1}{n\kappa_{n-\nu}}\binom{n}{\nu} W_{n-\nu}\quad\mathrm{for\ }\nu\leq n-1\;.
\]
A similar reindexing and rescaling applies to the Minkowski tensors defined in Chapter~\ref{chap-02}.
For $r,s\in \N_0$ we define following~\cite{McMullen1997, SchneiderWeil2008}
\[
W_0^{r,0}(K)=\int_K x^r \mathrm{d}x, 
\]
and 
\[
W_\nu^{r,s}(K)=\frac{n\kappa_{n-\nu}}{\binom{n}{\nu}}\int_{\R^n\times S^{n-1}} x^ru^s \Lambda_{n-\nu}(K;\mathrm{d}(x,u)) \quad\mathrm{for\ }n\geq \nu \geq 1 \;, 
\]
where $x^r$ and $u^s$ are symmetric tensor products, and $x^ru^s$ is the symmetric tensor product of $x^r$ and $u^s$,
and $\Lambda_\nu(K;\nularg)$ is  the $\nu$th support measure of $K$; see Sec.~\ref{01-sec3} in Chapter \ref{chap-01}.
Comparing this with \eqref{02-1.3}, \eqref{02-1.5}, and \eqref{02-1.6}, we see that 
\[
\Phi_n^{r,0}(K)= \frac1{r!} W_0^{r,0}(K),
\]
and 
\[
  \Phi_\nu^{r,s}(K)= \frac{\omega_{n-\nu}}{r!s!\omega_{n-\nu+s}\omega_{\nu}}\binom{n-1}{\nu-1}W_{n-\nu}^{r,s}(K) \quad\mathrm{for\ }n\geq \nu \geq 1 \;.
\]
\fi

In this work, we concentrate on the translation invariant Minkowski tensors $W_{\nu}^{0,2}$.
For a sufficiently smooth three-dimensional compact convex set $K$, these tensors of rank two can be represented by symmetric matrices 
\begin{align}
  W_1^{0,2} &= \int_{\partial K} \left(
  \begin{array}{c c}
    n_x^2 & n_xn_y\\
    n_xn_y & n_y^2
  \end{array}\right) \text{d}A,\\
  \intertext{and}
  W_2^{0,2} &= \int_{\partial K} \frac{\kappa_1+\kappa_2}{2} \cdot \left(
  \begin{array}{c c}
    n_x^2 & n_xn_y\\
    n_xn_y & n_y^2
  \end{array}\right) \text{d}A,
  \label{eq:def-wnu02}
\end{align}
where $\vec{n}=(n_x, n_y)$ describes the normal vector at the boundary
of $K$; $\kappa_1$ and $\kappa_2$ are the principle curvatures on
$\partial K$, and $(\kappa_1+\kappa_2)/{2}$ is the mean curvature.

\subsection{Geometrical interpretation}

In three dimensions, the Minkowski functionals are proportional to either the volume, the surface area, the integrated mean curvature, or the Euler characteristic.
The latter is a topological constant, which measures in a certain way connectivity.
For a single cell without holes, the Euler characteristic is trivially equal to unity.

The Minkowski functionals of a domain $K$ can be expressed by integrals over $K$ or over its boundary $\partial K$.
These scalar measures are naturally generalized to the Minkowski tensors by including an integral over the tensor products of the position vector $\vec{r}$ and the surface normal vector $\vec{n}$.
In other words, they are the moment tensors of the position or normal vectors.
The tensors using the position vector are closely related to tensors of inertia where the mass is located in the region of integration and possibly weighted by the curvature.
$W_0^{2,0}$ contains the information of the tensor of inertia of the solid object, $W_1^{2,0}$ of a hollow object where the mass is located in the shell.
For the example of polytopes in three dimensions, $W_2^{2,0}$ and $W_3^{2,0}$ are related to the tensor of inertia if the mass is distributed on the edges or vertices but weighted with the opening angles.
The tensor $W_1^{0,2}$ is proportional to the moment or covariance tensor of the distribution of normal vectors, in other words, of the orientation of the surface;
$W_2^{0,2}$ is proportional to the according moment tensor weighted by the curvature distribution.
In contrast to the tensors that are related to the tensors of inertia, the moment tensors of the normal distributions are translation invariant.

The Minkowski tensors allow for a systematic analysis of anisotropy w.r.t.  different geometrical aspects, like the distribution of volume or of the orientation of the surface.
While a domain $K$ might be perfectly isotropic w.r.t. one of these properties, it can be strongly anisotropic w.r.t. another property.
For example, a homogeneous random two-phase medium, like a stationary Boolean model, has an isotropic distribution of the volume; hence, the tensors $W_{\nu}^{2,0}$ are isotropic~\cite{HoerrmannEtAl2014, SchroederTurketal2010AdvMater}.
However, it can be strongly anisotropic w.r.t.  the distribution of the normal vectors on the interface, which is detected by $W_{1}^{0,2}$~\cite{HoerrmannEtAl2014, SchroederTurketal2010AdvMater}.

Different normalizations are used in different fields of research.
In this article, we use for the Minkowski functionals $W_{\nu}$ and tensors $W_{\nu}^{r,s}$ a normalization such that $W_{0}$ is the volume and $W_1$ is the surface area.
The normalizations of the Minkowski tensors are then defined accordingly~\cite{Klatt2016}.

Here, the tensor $W_{3}^{0,2}$ is not of interest because it is proportional to the unit tensor times the Euler characteristic (see detailed discussion in the conclusion section of~\cite{Mickel:2012}).
The latter is, as mentioned above, for a compact, convex cell always equal to one.

\subsection{Shape indices}

Even in a statistically isotropic random tessellation, a typical cell is usually geometrically anisotropic.
There is no globally preferred orientation.
However, given a specific realization of a single cell, the distribution of its normal vectors will be most likely non-uniform.
This geometric anisotropy of each single cell is here characterized by the Minkowski tensors.
Each tensor $W_{\nu}^{0,2}$ contains both information about the preferred direction and the amplitude or degree of the anisotropy.
The latter can conveniently be described by a scalar anisotropy index: the ratio of the smallest to the largest eigenvalue $\beta_{\nu}^{0,2}$ of the Minkowski tensor $W_{\nu}^{0,2}$~\cite{SchroederTurketal:2010jom, SchroederTurketal2010AdvMater}.
It describes the ``degree of anisotropy'' that is captured by the Minkowski tensor $W_{\nu}^{0,2}$.
(In physical literature, such shape indices are sometimes referred to as ``shape measures.''
However, this does not refer to a measure in the mathematical sense.) For example, if $\lambda_1$ and $\lambda_3$ are the smallest and largest eigenvalues of $W_1^{0,2}$, respectively, the anisotropy index is given by:
\begin{align}
  \beta_{1}^{0,2} := \frac{\lambda_1}{\lambda_3}.
  \label{eq:beta102def}
\end{align}
Smaller values indicate stronger anisotropy.
For a ball, $\beta_{1}^{0,2}=1$.
A cube also appears perfectly isotropic to a second rank tensor.
Therefore, $\beta_{1}^{0,2}=1$ for a cube.
For a cuboid, the index $\beta_{1}^{0,2}$ is equal to the ratio of the surface areas of the smallest and largest faces.

Alternative scalar anisotropy indices $q_r$ can also be derived for Minkowski tensors of arbitrary rank $r$~\cite{Kapfer2011, MickelEtAl2013}.
The index $q_2$, for example, is equivalent to a weighted bond orientational order parameter~\cite{MickelEtAl2013}.
For two-dimensions, see~\cite{Klatt2016}.

Given a sample of $M$ cells, we determine for each cell
$C_m$ (with $m\in\{1,\ldots M\}$) the Minkowski functionals
$W_{\nu}(C_m)$ and the anisotropy indices $\beta_{\nu}^{0,2}(C_m)$.
We then estimate the corresponding mean values for the typical cell by the sample
means:
\begin{align}
  \begin{aligned}
  \langle{}{W}_{\nu}{}\rangle&:=\frac{1}{M}\sum_{m=1}^{M}W_{\nu}(C_m),\\
  \langle{}{\beta}_{\nu}^{0,2}{}\rangle&:=\frac{1}{M}\sum_{m=1}^{M}{\beta}_{\nu}^{0,2}(C_m).
  \end{aligned}
  \label{eq:sample-mean}
\end{align}
Put differently, the shape index $W_{\nu}$ (or $\beta_{\nu}^{0,2}$) is
evaluated for each cell separately, and the arithmetic average is the
estimator for the expectation.
These estimators can be justified by the ergodicity properties of the underlying tessellations.
Similarly, the sample variance over all cells provides an estimate for
the variance:
\begin{align}
  s^2_{W_{\nu}} &:= \frac{1}{M-1}\sum_{m=1}^{M}(W_{\nu}(C_m)-\langle{}{W}_{\nu}{}\rangle)^2,\\
  s^2_{{\beta}_{\nu}^{0,2}} &:= \frac{1}{M-1}\sum_{m=1}^{M}({\beta}_{\nu}^{0,2}(C_m)-\langle{}{\beta}_{\nu}^{0,2}{}\rangle)^2.
\end{align}
Based on the sample mean and sample variance, we define normalized
Minkowski functionals and normalized anisotropy indices:
\begin{align}
  \label{eq:normalizedWnu}
  \hat{W}_{\nu} := \frac{{W}_{\nu} -
  \langle{}{W}_{\nu}{}\rangle}{s_{{W}_{\nu}}},\\
  \label{eq:normalizedBeta}
  \hat{\beta}_{\nu}^{0,2} := \frac{{\beta}_{\nu}^{0,2} - \langle{}{\beta}_{\nu}^{0,2}{}\rangle}{s_{{\beta}_{\nu}^{0,2}}}\;.
\end{align}

For these normalized or rescaled shape indices, we also determine estimated probability density functions (EPDFs) $f$,
that means empirical histograms weighted by the total number of samples and the bin width.
(A bin is a range of values for which the frequency of occurrence is determined.)
In other words, the EPDF is a relative frequency histogram weighted by the size of each bin.

In Section~\ref{sec:map}, we use another shape index
\begin{align}
  Q := 36\pi\cdot\langle V \rangle^2/\langle A \rangle^3,
  \label{eq:Qdef}
\end{align}
where $V(=W_0)$ is the volume of the cell, $A(\propto W_1)$ its surface area, and $\langle . \rangle$ denotes the sample mean as defined in Eq.~\eqref{eq:sample-mean}. 
The mean volume is rescaled by the mean surface area, so that the resulting index has no unit.
The prefactor $36\pi$ is chosen such that the upper bound (which is given by the perfectly isotropic sphere) is unity.

$Q$ can be interpreted as an isoperimetric ratio of the empirical average cell volume and area.
However, the ratio of mean values is in general not equal to the mean value of the corresponding ratio of volume and surface area.
Therefore, $Q$ can be quite different from the mean isoperimetric ratio (or quotient) of the typical cell.

For a single body $K$, the isoperimetric ratio is defined as $Q_s(K):=36\pi\cdot V(K)^2/A(K)^3$,
where $V(K)$ and $A(K)$ are the volume or surface area of $K$.
It characterizes to some extent the deviation from a spherical shape: only for a sphere $Q_s=1$, for all other bodies $Q_s < 1$.
The isoperimetric quotient is, e.\,g., used to describe equilibrium phases of packings of hard convex polyhedra.
It can be correlated to the isoperimetric ratio and the coordination number in a map of phases~\cite{Manoharan2015}
similar to our map of average anisotropies in the following Section~\ref{sec:map}.

\begin{figure}[p]
  \centering
  \includegraphics[width=0.9\textwidth]{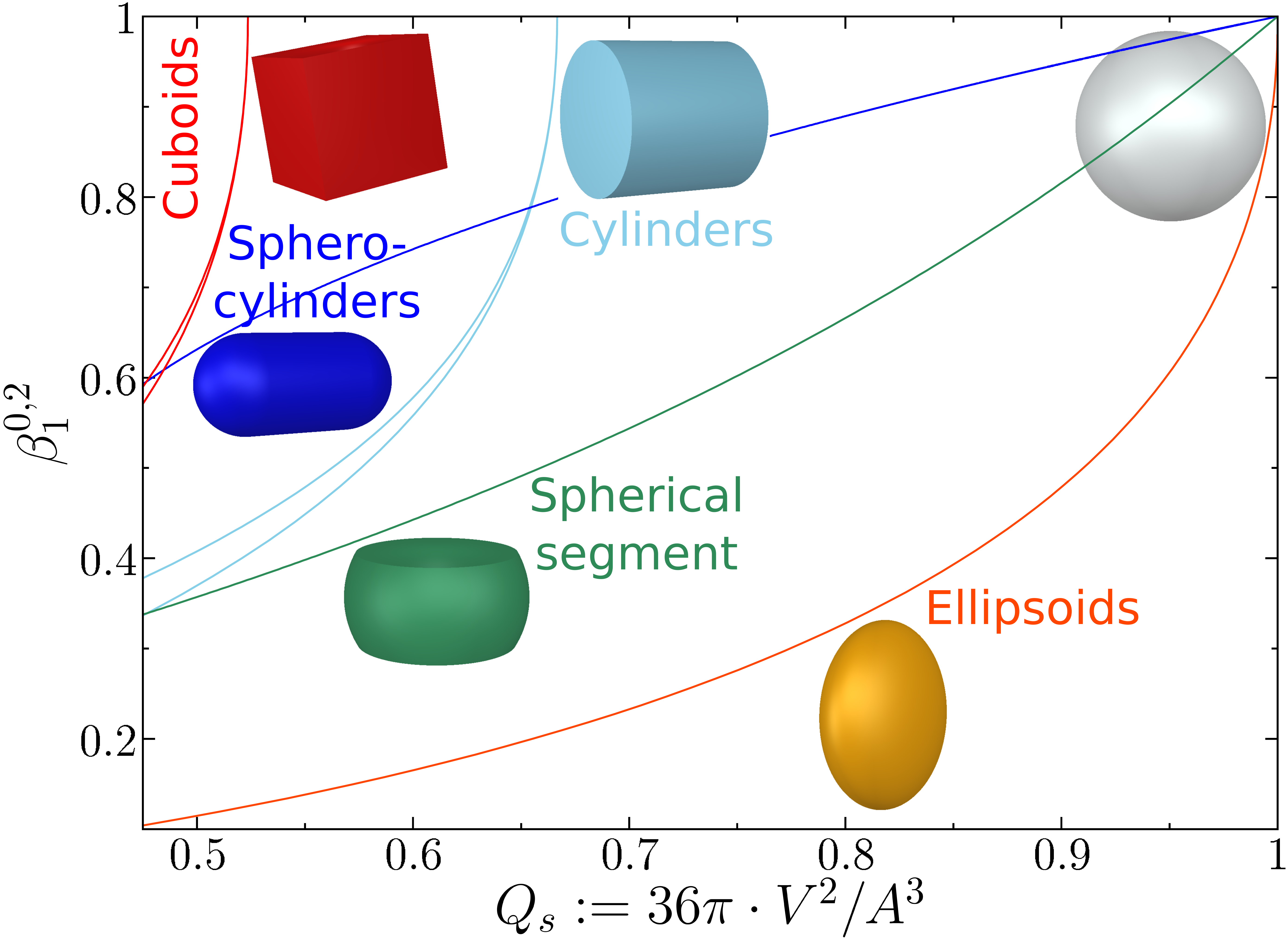}
  \caption{Anisotropy indices for exemplary single bodies.
  The curves show the anisotropy indices for a family of objects with different aspect ratios.}
\label{fig_single-solids_map}
\end{figure}

In Fig.~\ref{fig_single-solids_map}, the anistropy index $\beta_{1}^{0,2}$ from Eq.~\eqref{eq:beta102def} is compared for some exemplary single bodies to the isoperimetric ratio $Q_s$.
Only for a sphere, are both $\beta_{1}^{0,2}=1$ and $Q_s=1$.
For a cube, $\beta_{1}^{0,2}=1$ but $Q_s = \pi/6 \approx 0.52$.
The two (red) lines for cuboids in Fig.~\ref{fig_single-solids_map} correspond to either an elongation or a contraction of one of the sidelengths of a cube.
Similarly, the two (cyan) lines show the anisotropy indices of cylinders with a varying ratio of its height to its diameter, which can be larger or smaller than unity.
If the ratio is exactly equal to one, $\beta_{1}^{0,2}=1$.
If spherical caps are added to the cylinder, it is called a sphero-cylinder.
In the limit of a vanishing ratio of height to diameter, it becomes a sphere.
The spherical segments in Fig.~\ref{fig_single-solids_map} are defined by cutting a sphere with two parallel hyperplanes.
More precisely, they are the intersections of the unit sphere with $[-1,1]\times[-1,1]\times[-h,h]$.
The (green) line represents different values of this height $h$.
Finally, Figure~\ref{fig_single-solids_map} plots the anisotropy indices of prolate ellipsoids (where the two smaller principal axes are equally long) with different aspect ratios (orange line).


\section{Averages and map of mean anisotropy indices}
\label{sec:map}

First, we compare for the tessellations discussed in Section~\ref{sec:pp_def} the mean values of the Minkowski functionals and tensors of the typical cells.

\begin{figure}[p]
  \centering
  \includegraphics[width=0.9\textwidth]{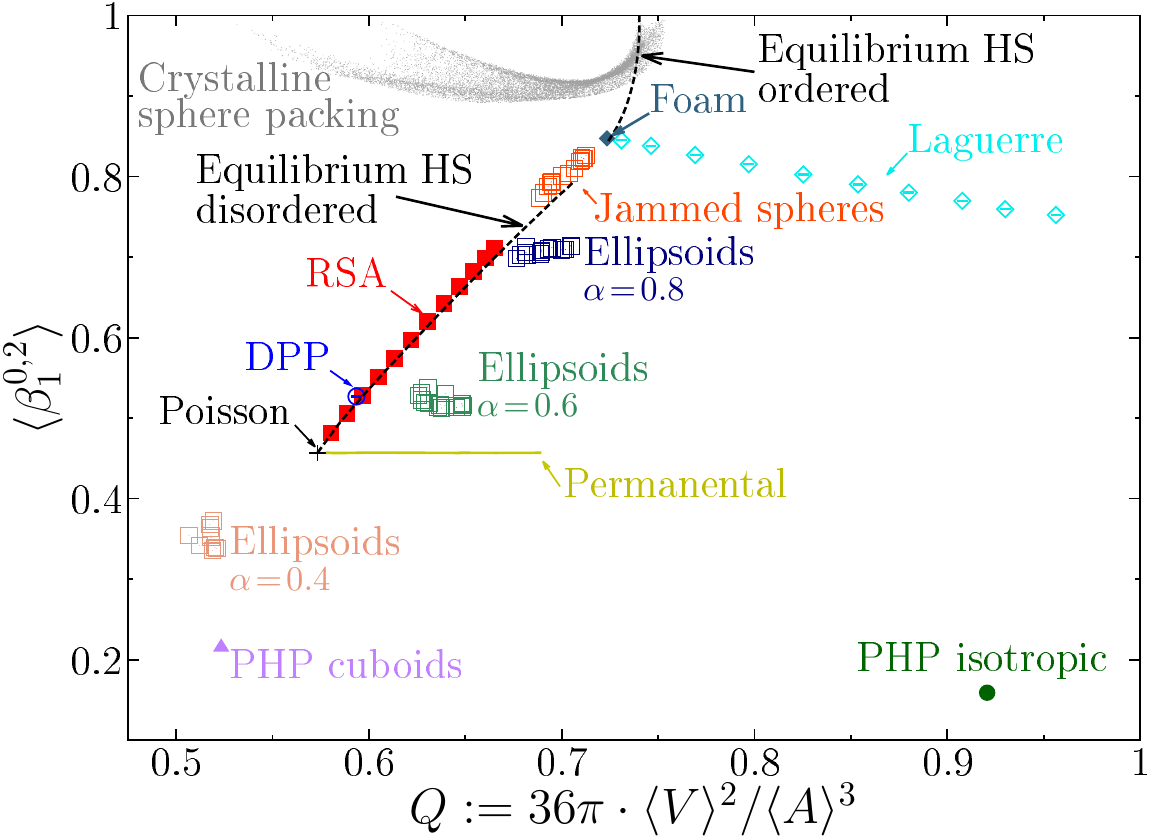}
  \caption{Map of local anisotropy for various tessellations ranging
    from Poisson and equilibrium hard sphere (HS) Voronoi tessellations, over PHP tessellations to
    cells in crystalline sphere packings. For each tessellation, the mean
    anisotropy index $\langle\beta_1^{0,2}\rangle$, see Eq.~\protect\eqref{eq:beta102def}, and
    the parameter $Q$ from Eq.~\protect\eqref{eq:Qdef} are shown.
    For the jammed ellipsoid packings, experimental data is shown for
    different aspect ratios $\alpha$. The map of anisotropy reveals
    structural differences between the models, but also how they are
    related to each other.}
\label{fig_map}
\end{figure}

These averages allow us to construct a ``map of anisotropy''~\cite{Klatt2016}, see Fig.~\ref{fig_map}.
Each tessellation is represented by its mean anisotropy index $\langle \beta_1^{0,2} \rangle$, see Eq.~\eqref{eq:beta102def}, and the parameter $Q$ from Eq.~\eqref{eq:Qdef}.
The map of anisotropy provides a kind of classification of the tessellation models based on their anisotropy, which can, e.\,g., help to choose the appropriate model for applications. 

It also reveals how some of the models can be related to each other.
For example, the systems of equilibrium hard spheres at different global packing fractions connect,
in the two limits of vanishing or maximal packing fraction, the uncorrelated Poisson point process $\left(+\right)$
and the perfectly ordered close packing on a lattice (gray dots).
The anisotropy indices for the equilibrium hard spheres are represented by two dashed (black) lines.
The curve separates into an ordered and a disordered branch because of the disorder-to-order phase transition, see Section~\ref{sec:equihs}.
(For clarity of presentation, we connect single data points close to each other by a dashed line.)

At the densest global packing fraction $\phi_{\mathrm{max}}=\pi/\sqrt{18}\approx 0.74$~\cite{Hales2005},
the spheres form a face-centered-cubic crystal or one of its stacking variants.
Due to the symmetry of these crystals, the single cells are perfectly isotropic w.r.t. the anisotropy index $\beta_1^{0,2}$ of the second-rank Minkowski tensor.
The cells in the other crystalline sphere packings (gray dots) are, as expected, also either isotropic or only slightly anisotropic w.r.t. $\beta_1^{0,2}$.

For the RSA process $\left(\blacksquare\right)$, different points in the figure correspond to different global packing fractions.
The Poisson point process $\left(+\right)$ is quite irregular, and its Voronoi cells are therefore rather anisotropic.
In the dilute limit the RSA process, like the equilibrium hard spheres, have a structure similar to a Poisson point process
(because interactions between the particles are unlikely).
However, the denser the hard-particle systems get, the more correlated and ordered they become.
Therefore, the typical cells become more isotropic w.r.t. both $\beta_1^{0,2}$ and the ratio $Q$.
Put differently, an increasing packing fraction decreases the anisotropy in the hard-particle systems ($\langle\beta_1^{0,2}\rangle$ gets closer to unity).

In Fig.~\ref{fig_map}, the mean anisotropy indices of the DPP $\left(\circ\right)$ are similar to that of an RSA pattern $\left(\blacksquare\right)$.
However, the mean values for the packings of jammed ellipsoids $\left(\square\right)$ are distinctly different.
For ellipsoids with an aspect ratio $\alpha=0.4$, the corresponding set Voronoi cells are on average more anisotropic than Poisson Voronoi cells.

The different points for the Laguerre tessellations $\left(\diamond\right)$ correspond to different polydispersities of the underlying hard sphere packings, that is different degrees of variation in the sphere volumes.
A stronger polydispersity decreases $\langle\beta_1^{0,2}\rangle$, i.\,e., the typical cell is more anisotropic.
At the same time, however, the ratio $Q$ increases, for which we can provide a heuristic argument.
The estimate $\langle\beta_1^{0,2}\rangle$ of the mean value for a typical cell is dominated by a vast number of small anisotropic cells.
The same would apply to the mean isoperimetric ratio $\langle Q_s\rangle$ of a typical cell.
However, $Q$ is a ratio of mean values.
Larges cells strongly influence the mean values of volume and surface area, in the sense, that many small cells can be approximated by zero volume and surface area compared to cells that are by orders of magnitude larger.
Therefore, the ratio $Q$ will be closer to the isoperimetric ratio of large cells which are on average more isotropic.
Because $\langle \beta_1^{0,2} \rangle$ is the mean anisotropy index of the typical cell, a comparison to $Q$, which is the ratio of mean values, can distinguish more systems than a comparison to $\langle Q_s\rangle$, which is the mean isoperimetric ratio of the typical cell.
In the latter case, two mean indices of the typical cell are compared, in contrast to Fig.~\ref{fig_map} which is based on a mean value $\langle \beta_1^{0,2} \rangle$ and a ratio $Q$ of mean values.

For a similar reason, the isotropic PHP tessellations $\left(\bullet\right)$ exhibit a small anisotropy index $\langle\beta_1^{0,2}\rangle$
but a large ratio $Q$.
In Section~\ref{sec:PHP-STIT-local-beta}, we discuss in more detail the average anisotropy for differently large cells.
The average anisotropy of a typical cell in a statistically isotropic PHP tessellation is more
anisotropic than the average anisotropy of a typical cell in a PHP tessellation with cuboids as cells $\left(\blacktriangle\right)$.

The data for the permanental point process is represented by the solid (yellow) line which connects the single data points.
Because a single realization of a permanental point process is here the outcome of an inhomogeneous PPP, the pattern can locally be very similar to a homogeneous PPP.
Therefore, $\langle \beta_1^{0,2} \rangle$ of the typical Voronoi cell is for the models that we have simulated similar to $\langle \beta_1^{0,2} \rangle$ of a typical Poisson Voronoi cell.
However, with increasingly anisotropic underlying Gaussian random fields, the index $Q$ becomes larger (possibly because of a stronger polydispersity of the cells like for the Laguerre tessellations).

For the monodisperse foam $\left(\vardiamond\right)$, the jammed sphere packings $\left(\square\right)$, and the Laguerre tessellations in the monodisperse limit $\left(\diamond\right)$, the anisotropy takes on very similar values.
The tessellations are all related in that they are at least based on the Voronoi tessellation of a rather dense and disordered packing of hard spheres, see Section~\ref{sec:pp_def}.
(Note that an equilibrium hard-sphere liquid at the same global packing fraction is significantly more regular than these disordered packings.)
However, there are also distinctive geometric differences between these systems, which this coarse analysis cannot capture. For example, the faces of a foam cell are curved~\cite{Kraynik2006}.

Moreover, the anisotropy of the Voronoi tessellations for the RSA and equilibrium hard-sphere systems as well as for the DPP, are at least for the range of parameters chosen here very similar.
All three point processes have in common that they are purely repulsive.


\section{Shape distribution functions}
\label{sec:distributions}

Going beyond the average, we consider the full probability density function of the Minkowski functionals of the typical cell as well as of the anisotropy index.

\begin{figure}[t]
  \centering
  \includegraphics[width=\textwidth]{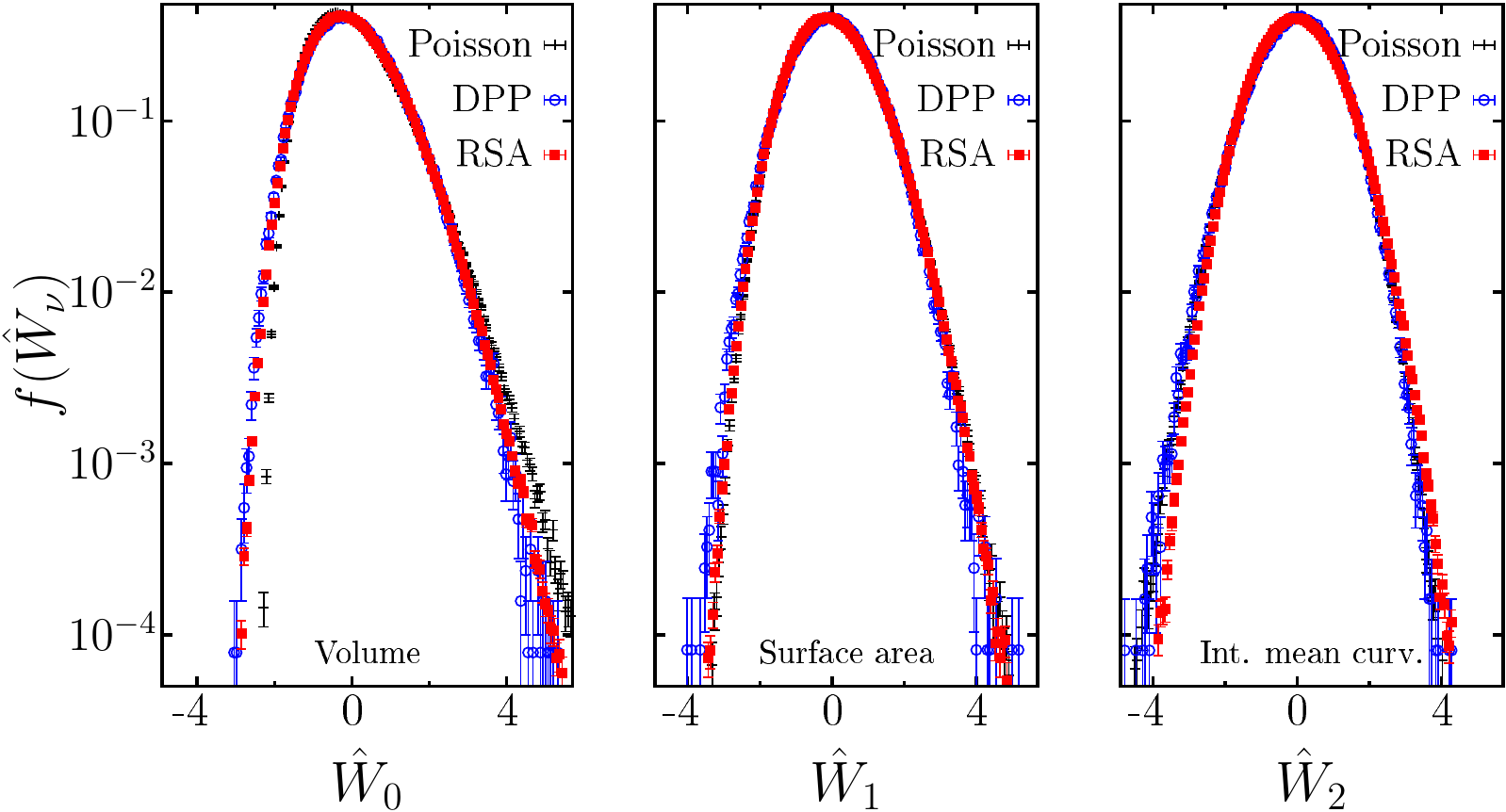}
  \caption{Estimated probability density functions $f$ of the normalized
    Minkowski functionals $\hat{W}_{\nu}$, see
    Eq.~\protect\eqref{eq:normalizedWnu}, for cells in Voronoi tessellations of three
  different point processes: Poisson point process (see
  Section~\ref{subPoisson}), DPP (see
  Section~\ref{sec:dpp}), and RSA (see Section~\ref{sec:rsa}).}
  \label{fig:epdf-Voronoi}
\end{figure}

\begin{figure}[b]
  \centering
  \includegraphics[width=\textwidth]{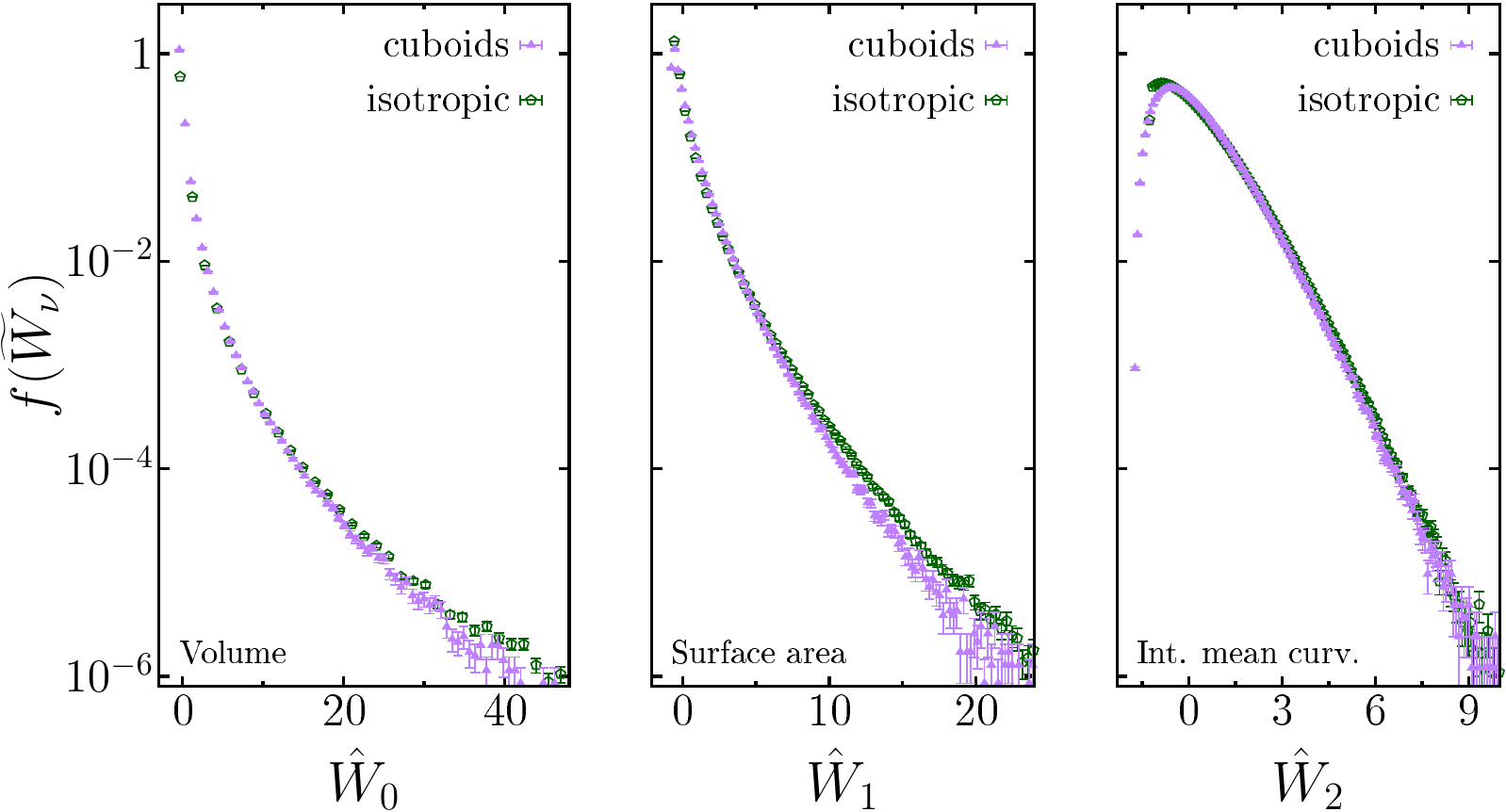}
  \caption{Estimated probability density functions $f$ of the normalized
    Minkowski functionals $\hat{W}_{\nu}$, see
    Eq.~\protect\eqref{eq:normalizedWnu}, for cells in PHP tessellations with
    different orientation distributions of the hyperplanes (see
    Section~\ref{sec:Poisshyp}).}
  \label{fig:epdf-PHP}
\end{figure}

From the data (described in Section~\ref{sec:pp_def}), we determine the EPDFs of the normalized Minkowski functionals and of the normalized anisotropy indices, which are derived from the Minkowski tensors.
Figures~\ref{fig:epdf-Voronoi} and \ref{fig:epdf-PHP} show the resulting curves.


How sensitive is this description of the qualitative  behavior of
several features of the local structure?  Figures~\ref{fig:epdf-Voronoi}
and \ref{fig:epdf-PHP} clearly show that the probability density
functions of the Minkowski functionals are distinctly qualitatively
different for the Voronoi or hyperplane tessellations.
At least for the here simulated models of PHP (or equivalently STIT) tessellations, the probability density function of the volume appears to be monotonically decreasing at least for a large range of volumes.
In contrast to this, the distributions of the Voronoi cell volumes show a clear maximum close to the mean values.

In the latter case, the distributions are for several point processes well-known to be in good agreement with (generalized) Gamma distributions~\cite{Pineda2004, AsteMatteo2008, KapferEtAl2010, LazarEtAl2013}.
However, statistically significant deviations have been found for jammed particle packings~\cite{KlattTorquato2014}.

The occurrence of Gamma distributions in tessellations driven by Poisson processes of points or flats was observed in the seminal mathematical work~\cite{Miles71,MZ960}.
Even for Poisson point processes on very general spaces the intensity measure of certain random sets is conditionally Gamma-distributed~\cite{Z99}.
A systematic and unifying explanation of these phenomena in a Euclidean setting was given in~\cite{BaumstarkLast09}; see also~\cite{BaumstarkLast07}.
One of the results in~\cite{MZ960,BaumstarkLast09} ascertains (for $n=3$) that the distribution of the $(n-2)$-nd Minkowski functional of the typical cell of a statistically isotropic Poisson hyperplane tessellation is conditionally Gamma-distributed given the number $m$ of neighbors.
The shape parameter of this Gamma distribution is just $m-n$.
Hence, the unconditional distribution is a mixture of Gamma distributions (with the same scale parameter) with respect to the distribution of the number of neighbors of the typical cell.

In a Poisson hyperplane tessellation, the number of neighbors of a cell coincides with the number of $(n-1)$-dimensional facets.
Note that this is in general not the case in a STIT tessellation.
(However, the distribution of the number of $(n-1)$-facets of the typical cell and the distributions of the Minkowski functionals are the same as for the corresponding PHP tessellation.)

\begin{figure}[t]
  \centering
  \subfigure[][]{\includegraphics[height=5.5cm]{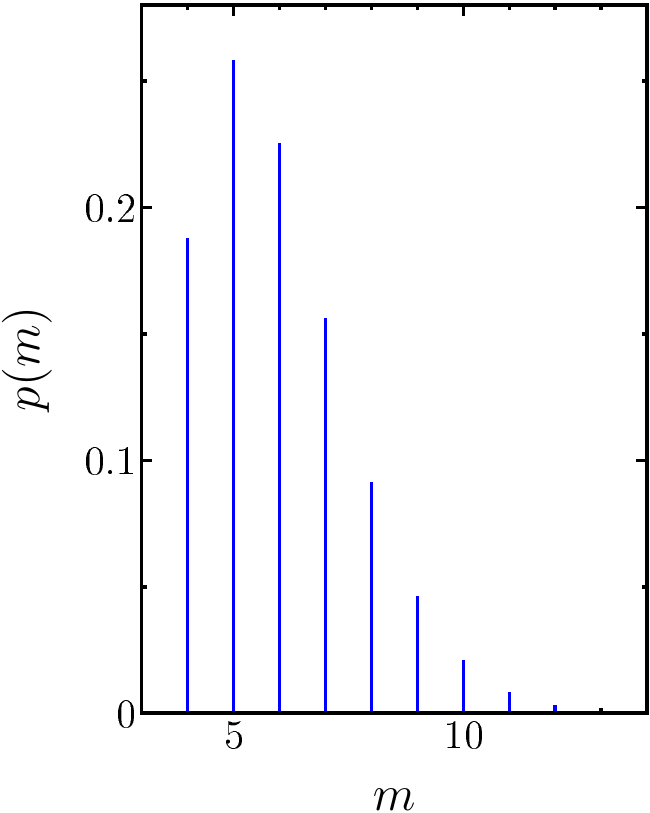}}
  \hfill
  \subfigure[][]{\includegraphics[height=5.5cm]{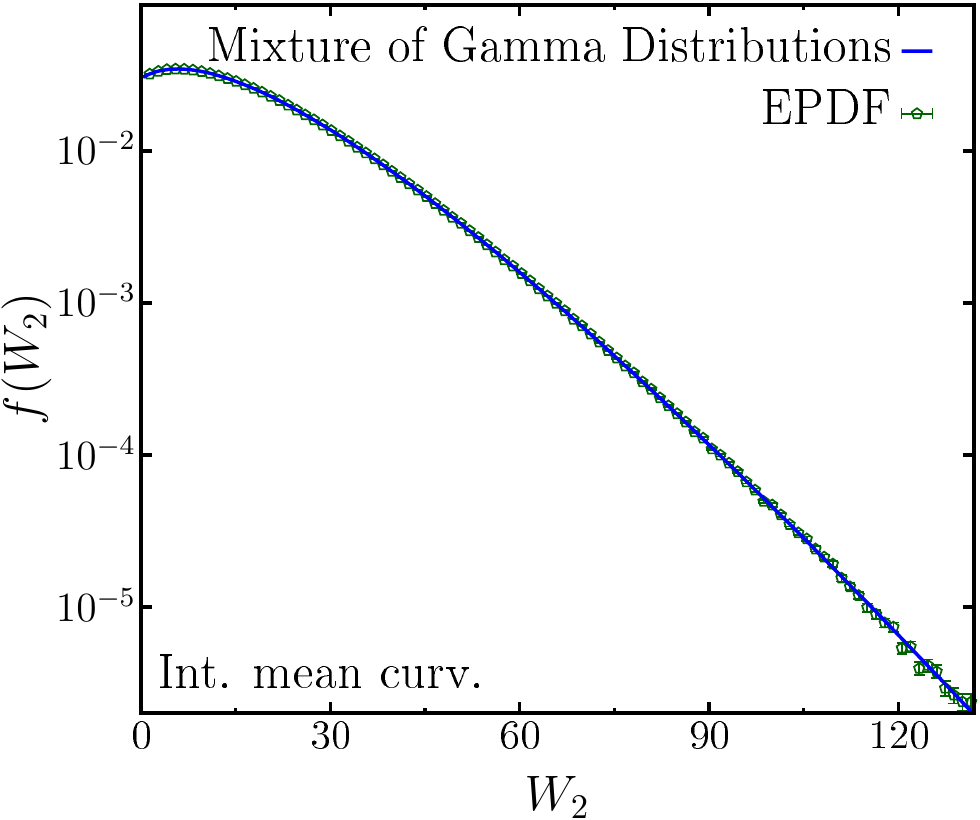}}
  \caption{Relation between the distribution of the number of faces and
    the Minkowski functional $W_2$ in an isotropic PHP tessellation:
    (a) the empirical probability mass function $p$ of the number $m$
    of two-dimensional facets; (b) the EPDF $f(W_2)$ of the integrated
    mean curvature of a cell (marks) is in very good agreement with the mixture of Gamma
distributions given in Eq.~\protect\eqref{eq:eq-W2} and using the empirical
probability mass function $p$ in (a).}
  \label{fig:p-m-W2}
\end{figure}

According to~\cite{BaumstarkLast09}, in both cases the probability density function of the $(n-2)$-nd Minkowski functional of the typical cell
can be expressed by the probability mass function $p$ of the number of $(n-1)$-dimensional facets:
\begin{align}
  f(W_{n-1}) = \frac{1}{2\kappa_{n-1}} \sum_{m=4}^{\infty} p(m) \cdot g(m-n,\gamma;\frac{W_{n-1}}{2\kappa_{n-1}})\;,
  \label{eq:eq-W2}
\end{align}
where $g(\alpha,\beta;x)=\beta^{\alpha}x^{\alpha-1}\text{e}^{-x\beta}/\Gamma(\alpha)$
is the probability density function of the Gamma distribution.
This is demonstrated in Fig.~\ref{fig:p-m-W2} (for $n=3$),
where the mixture of Gamma distributions given in Eq.~\eqref{eq:eq-W2} (the solid line in Fig.~\ref{fig:p-m-W2}~(b)) is in very good agreement with the EPDF of $W_2$.
In a three-dimensional stationary random hyperplane tessellation (with finite intensity), the average number of facets of the typical cell is $n_{3,2}=6$~\cite[Eq.~(10.35), p.  484]{SchneiderWeil2008},
which is in agreement with the sample mean $\langle m \rangle = 5.9994$ (where the standard error of the mean is $\approx 0.0004$).

Equation~\eqref{eq:eq-W2} shows that the information content of the number of facets and of the Minkowski functional $W_{n-1}$ is somehow related.
However, the latter is an additive, continuous measure where small changes in the positions and orientations of the hyperplanes correspond to only small changes in the Minkowski functionals.
This is in contrast to the number of facets which is a topological index, which can change distinctly and discontinuously for small variations of the positions and orientations of the hyperplanes.
Moreover, in experimental observations, the resolution of small facets might not be possible.
Therefore, the Minkowski functionals are more robust structure characteristics, which are also suitable for an analysis of noisy data sets.

The above described phenomenon extends to typical faces of lower dimensions~\cite{BaumstarkLast09}.
For other intrinsic volumes, there seems to be no mathematical argument supporting the occurrence of mixed Gamma distributions.

For the Poisson Voronoi tessellation, the situation is more complicated.
There is no obvious reason for the typical cell to have conditionally Gamma-distributed Minkowski functionals.
There is, however, the concept of the (typical) Voronoi flower, whose geometry is closely connected to the typical cell.
Given the number $m$ of neighbors of the typical cell, the volume of this flower has a Gamma distribution with shape parameter $m$~\cite{MZ960}.
Again this can be extended to flowers of typical faces of lower dimensions~\cite{BaumstarkLast07, BaumstarkLast09}.
There are some reasons to believe that similar results hold for other Minkowski functionals.

The agreement of the EPDFs of the normalized Minkowski functionals of Voronoi cells for very different point processes reveals some limitations of this univariate qualitative shape descriptors.
For physically different systems like the relatively long-ranged DPP, the uncorrelated PPP, or a non-equilibrium RSA system with only short-ranged interactions, the EPDFs are at least very similar.
For the systems studied here, there are small but statistically significant deviations only for the volume distribution of the Poisson Voronoi cells and for the distribution of the integrated mean curvature for the cells of an RSA process.
On the one hand, this reveals an interesting similarity in the local structure of Voronoi diagrams of random point processes.
On the other hand, this agreement for very different systems indicates that the EPDF of the normalized Minkowski functionals is not sensitive enough to detect the structural differences between these systems.
Note that the global structure differs distinctly for the four systems.
Such differences in the global structure of the Poisson point process, the equilibrium hard-sphere liquid, or a non-equilibrium jammed packing of hard spheres is, for example, discussed in detail in~\cite{KlattTorquato2014}.

\begin{figure}[t]
  \centering
  (a)
  \subfigure{\includegraphics[width=0.44\textwidth]{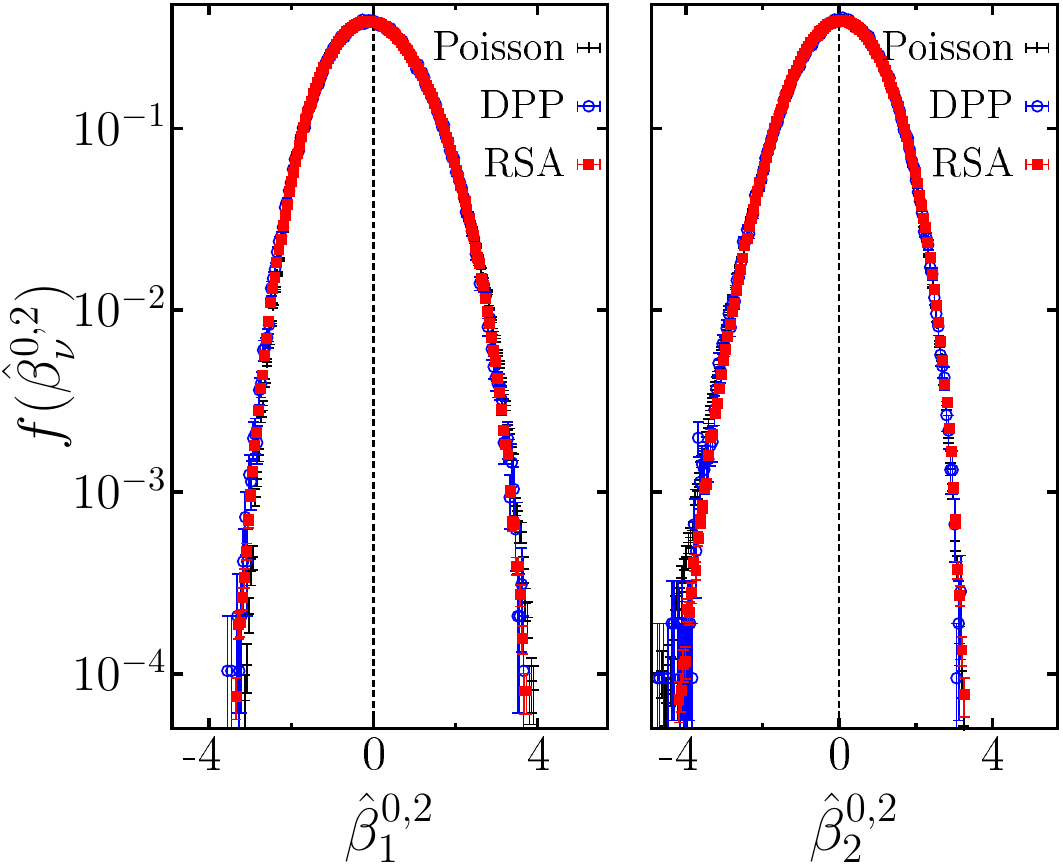}\label{fig:epdf-beta-Voronoi}}
  \hfill(b)
  \subfigure{\includegraphics[width=0.44\textwidth]{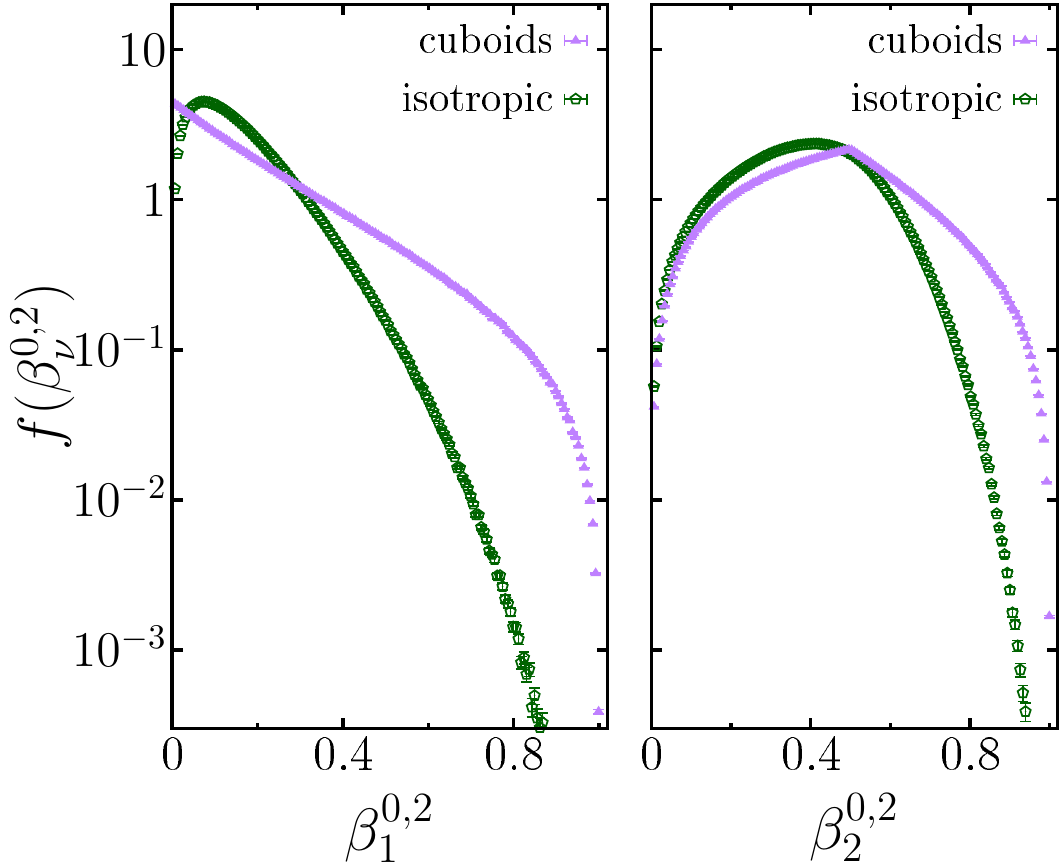}\label{fig:epdf-beta-PHP}}
  \caption{Estimated probability density functions $f$ (a) of the
    normalized Minkowski tensor anisotropy indices
    $\hat{\beta}_{\nu}^{0,2}$, see Eq.~\protect\eqref{eq:normalizedBeta}, for
    cells in Voronoi tessellations and (b) of the anisotropy indices
    ${\beta}_{\nu}^{0,2}$ for cells in PHP tessellations.}
  \label{fig:epdf-beta}
\end{figure}

Also the EPDFs of the anisotropy indices show a qualitatively different behavior for both PHP tessellations (and thus for the corresponding STIT tessellations) and the Voronoi tessellations.
However, for the three different Voronoi tessellations of a Poisson point process, the RSA process, or the DPP the same indifference of the curves appears as for the Minkowski functionals, see Fig.~\ref{fig:epdf-beta}.
This holds both for the anisotropy index $\beta_1^{0,2}$ of the surface tensor and for $\beta_2^{0,2}$ of the curvature tensor. 
An analysis based on a single characteristic is not able to resolve differences in the local structure.


\section{Local joint characterization via volume and anisotropy}
\label{sec:joint-characterization}

For a more sensitive characterization of the local structure, we need to take the relations between different characteristics into account.
By doing so, we can show that the local structure of these systems can indeed be qualitatively different although it seems indistinguishable in an analysis of single characteristics like in the Section above.
Moreover, this sensitive local analysis allows for intuitive geometric insights.

Following the analysis of \cite{SchallerEtAl2015epl}, we perform a conditional analysis based on the cell volume and consider the shape of small or large cells separately.
We quantify the anisotropy of the cells by the Minkowski tensor $W_1^{0,2}$ conditional on their volume.
In other words, we estimate the conditional expectation $\langle \beta_1^{0,2}\rangle_V$ of the anisotropy index $\beta_1^{0,2}$ as a function of the cell volume $V$.
More precisely, the condition is on the cell volume being in an interval $[V-\Delta V;V+\Delta V)$, i.\,e., we bin the cell volumes and then estimate $\langle \beta_1^{0,2}\rangle_V$ separately for each bin.

\subsection{Distinguishing local structures}

\begin{figure}[t]
  \centering
  \includegraphics[width=0.8\textwidth]{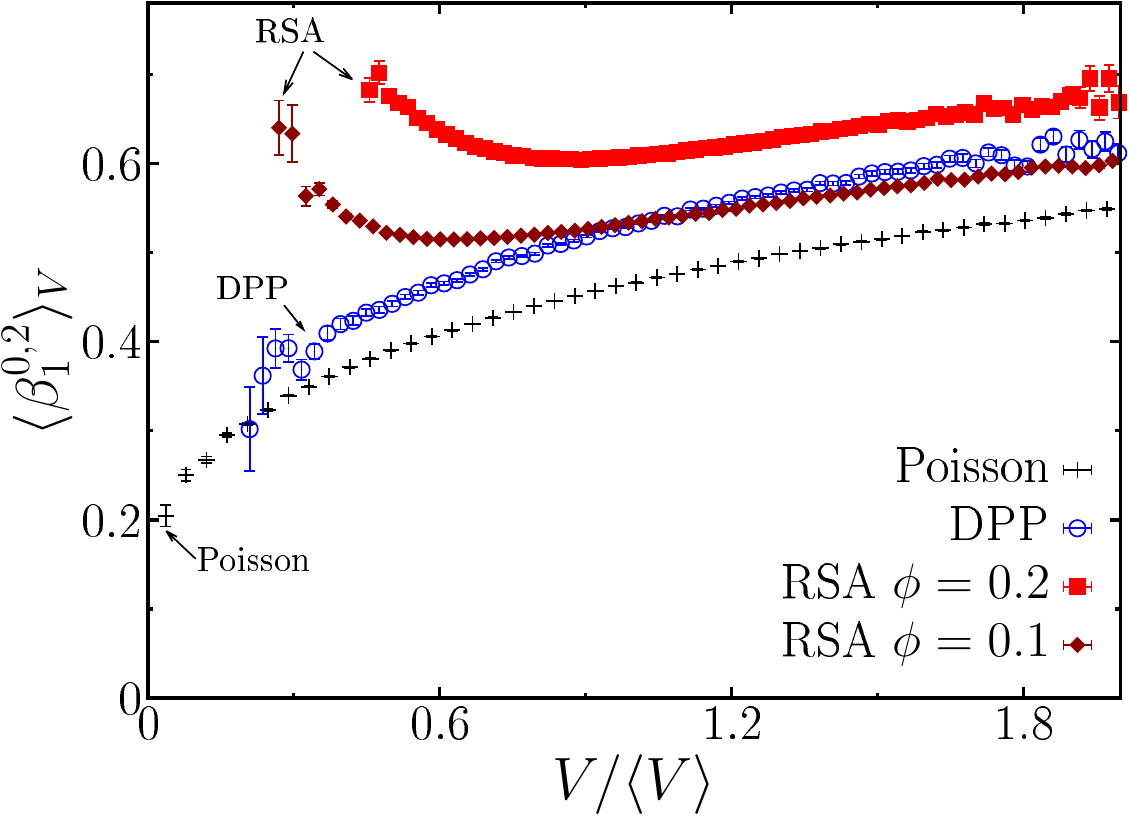}
  \caption{Anisotropy as a function of cell volume for Voronoi
    tessellations of a Poisson point process, a DPP, or RSA
    hard-sphere process with volume fractions $\phi=0.1$ or
    $0.2$. The combined analysis based on volume and anisotropy
    reveals the qualitatively different behavior of the different
    models.}
  \label{fig:v-beta-Poisson-DPP-RSA}
\end{figure}

Figure~\ref{fig:v-beta-Poisson-DPP-RSA} shows the resulting curves for the Voronoi tessellations of the PPP, RSA process, and DPP.
In contrast to the shape description by a single index in Section~\ref{sec:distributions}, the new analysis combining two shape indices can qualitatively distinguish the different point processes.

The Poisson points are non-interacting.
The hard spheres in the RSA process are rigid, i.\,e., perfectly repulsive at contact.
The points in the DPP can be interpreted as ``soft'' repulsive particles.
Although unlikely, the particles can get arbitrarily close to each other, which is in contrast to the RSA process.

The cells in the Poisson Voronoi tessellation are on average most anisotropic for all cell volumes.
The mean anisotropy index conditional on the cell volume is also (at least for a large range of cell volumes) smaller than for the Voronoi cells of the other systems.
The anisotropy index increases monotonically with increasing cell volumes, i.\,e., larger cells are on average more isotropic than smaller ones.
It is well known that the shape of a typical large cell in the Poisson Voronoi tessellation converges to a sphere in the limit of arbitrarily large cell volume~\cite{HugReitznerSchneider2004}.
Therefore, also the anisotropy parameter $\langle\beta_1^{0,2}\rangle_V$ must approach unity, i.\,e., perfect isotropy.

As expected, the RSA process of hard-spheres at different global packing fractions $\phi=0.1$ and $\phi=0.2$ have qualitatively similar curves in Fig.~\ref{fig:v-beta-Poisson-DPP-RSA}.
Because a hard-sphere system at finite packing fraction is more ordered than a Poisson point process,
it can be expected that a typical cell is more isotropic in the hard-sphere system.
This corresponds to an increase in the mean anisotropy index.
For large cell volumes, the anisotropy index also increases as a function of the cell volume.
However, in contrast to the uncorrelated Poisson point process, the curves for the RSA process exhibit a minimum.

For small cells, the trend reverses and the smaller the cell, the larger the anisotropy index gets.
The smallest possible Voronoi cell for a hard sphere packing is well-known to be a regular dodecahedron where the central sphere touches all faces.
This is the dual to an icosahedral arrangement of the neighboring spheres, which is the locally densest possible configuration with a maximum of 12 contacting neighbors.
(The volume of the smallest Voronoi cell for hard spheres divided by the mean Voronoi volume is given by $V/\langle V \rangle = 6/(\pi\sqrt{5/2+11/10\sqrt{5}})\cdot\phi\approx 0.1729\cdot\phi$ and rather small for the here chosen global packing fraction $\phi$.) In this limit, the anisotropy index converges to unity (because a regular dodecahedron appears perfectly isotropic w.r.t.  $\beta_{\nu}^{0,2}$).
Therefore, there is a minimum in the anisotropy index as a function of the cell volume.
The curve is qualitatively different from the corresponding curve for a Poisson Voronoi tessellation.

Moreover, it is also qualitatively different from the DPP.
Interestingly, the repulsive particles in the DPP show an intermediate behavior between the hard spheres and the non-interacting Poisson points.
For large cells, the anisotropy of the Voronoi cells of the DPP is comparable to those of the RSA process.
It is more regular than a PPP.
However, in contrast to the hard-sphere system, small cells get on average more anisotropic ($\langle\beta_1^{0,2}\rangle$ decreases).
These small cells are more similar to those in the irregular PPP.
A heuristic explanation of this behavior is that it is unlikely but possible that two points get close to each other.
However, it is then very unlikely that also a third point is located nearby.
Therefore, the two corresponding cells would, in this case, tend to be elongated because they are strongly restricted in the direction of the nearest neighbors.

Be reminded that the typical cells in a dilute hard-sphere gas or a weakly interacting DPP are very similar to those of a PPP, i.\,e., the mean values are very close to each other, see Fig.~\ref{fig_map}.
Moreover, even the empirically estimated full probability density functions are nearly indistinguishable, see Figs.~\ref{fig:epdf-Voronoi} and \ref{fig:epdf-beta-Voronoi}.
Analyzed by single characteristics alone, the local structure seemed at least qualitatively the same for the three different point processes.

In contrast to both the mean value and the univariate probability density functions, the  characterization based on both the volume and the anisotropy can clearly distinguish the Poisson Voronoi tessellation from the hard-sphere system.
The local structure is actually not the same, but rather the other shape descriptors were not sensitive enough to find a qualitatively different behavior.
In this case, our more sensitive local analysis can distinguish these different generating processes. 

\subsection{Equilibrium and non-equilibrium hard particles}

\begin{figure}[t]
  \centering
  \subfigure[][]{\includegraphics[height=4.2cm]{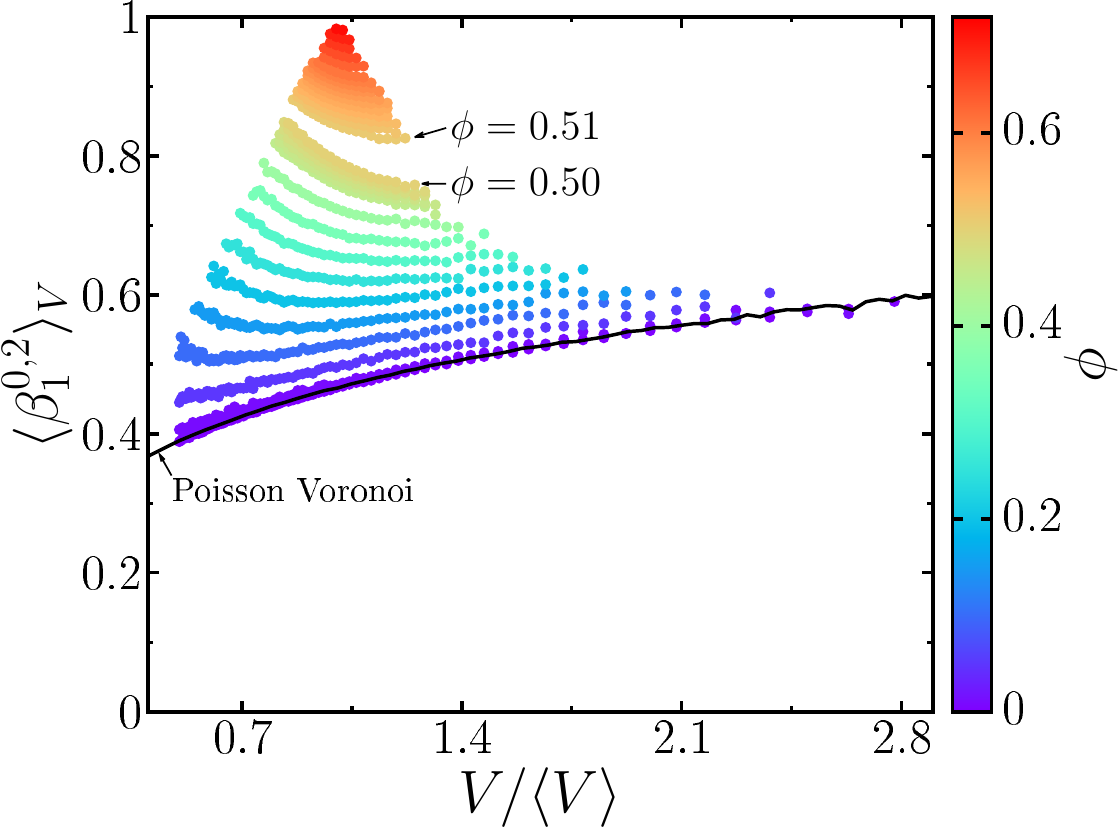}
  \label{fig:v-beta-equi}}
  \hfill
  \subfigure[][]{\includegraphics[height=4.2cm]{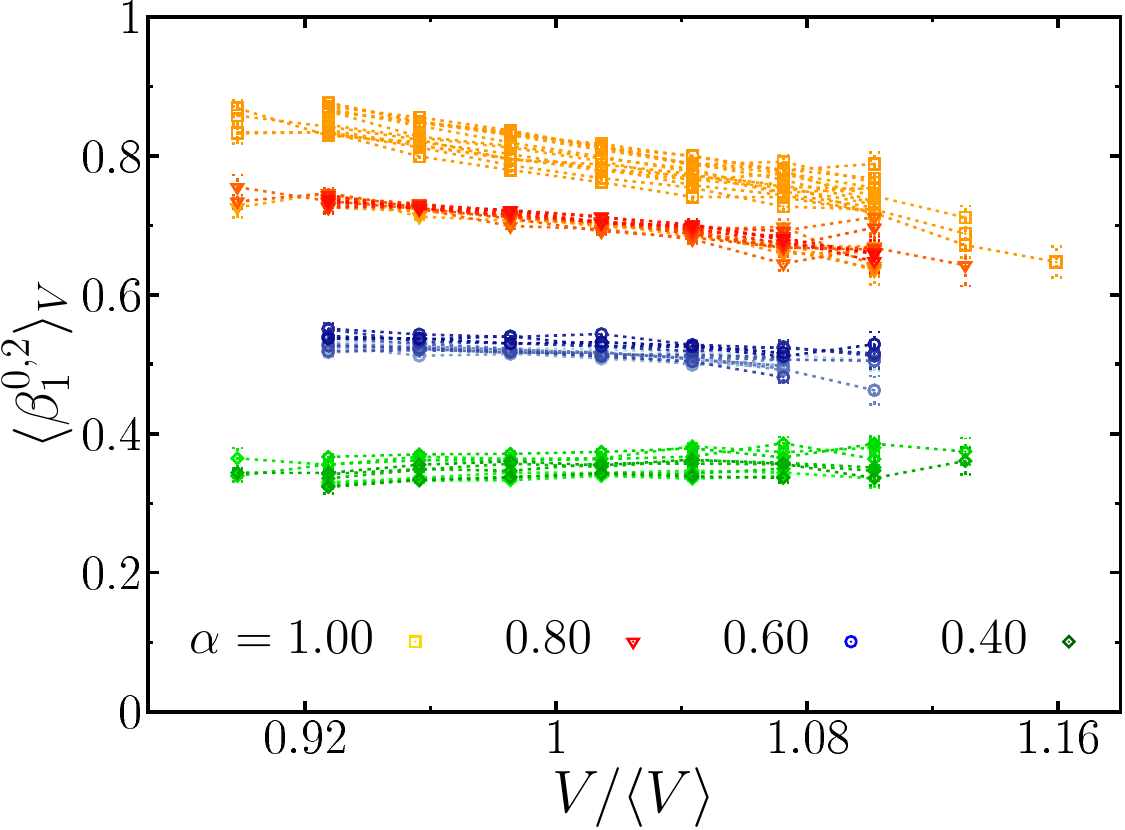}
  \label{fig:v-beta-jammed}}
  \caption{Anisotropy as a function of cell volume for hard-particle
    systems: (a) equilibrium hard-sphere liquids from the dilute limit
    (i.\,e., vanishing packing fraction) to nearly
    crystalline structures; the color scale indicates the varying
    packing fraction $\phi$; the black line corresponds to the Poisson
    point process (see also Fig.~\ref{fig:v-beta-Poisson-DPP-RSA}); the gap
    at roughly $\phi=0.5$ results from the freezing transition; (b)
    experimental data of random jammed ellipsoids for different aspect
    ratios $\alpha$ (indicated by different colors); the different
    curves with the same colors represent different experimental
    realizations with varying global packing fraction ($\phi = 0.54 - 0.68$)~\protect\cite{SchallerEtAl2015epl}.}
  \label{fig:v-beta-equi-jammed}
\end{figure}

Using this improved local shape analysis, Figure~\ref{fig:v-beta-equi-jammed} compares the local structure of the equilibrium hard-sphere liquids (a) and the non-equilibrium jammed ellipsoids (b).

In Fig.~\ref{fig:v-beta-equi}, the curve for the Poisson point process is extended to larger cell volumes.
As mentioned above, it must converge to unity.
However, the simulation reveals that it converges rather slowly.
Only the extremely dilute hard-sphere system are numerically difficult to distinguish from the PPP (because of the vanishing number of close neighbors and thus of small Voronoi cells).
The anisotropy as a function of the volume deviates for the equilibrium liquid already at relatively small global packing fractions from the uncorrelated PPP.

With increasing packing fraction, the range of observed Voronoi cell volumes shrinks because the configurations become more regular and thus the fluctuations in the Voronoi volume decrease.
In the limit of maximal packing fraction, only a single value of the volume is possible (which corresponds to the dual of the unit cell).

At a global packing fraction of about $\phi=0.5$, a gap in the curves of the anisotropy index is observed, see Fig.~\ref{fig:v-beta-equi}.
This is related to the solid-liquid hard-sphere phase transition.
Our samples are initially prepared in a crystalline state before equilibration.
Therefore, the transition occurs at the lower end of coexistence regime which is for an equilibrium hard-sphere system between $\phi \approx 0.494$ and 0.545 \cite{KapferEtAl2010}.

In contrast to the globally loose fluids where the system behaves like the PPP,
the anisotropy index in globally dense systems decreases monotonically as a function the cell volume.
This shows that in dense hard-sphere liquids the locally dense configurations are more ordered than the looser ones, and thus more isotropic.

For the jammed ellipsoid packings, the monotony of this functions changes with the aspect ratio of the ellipsoids.
Like in dense equilibrium hard-sphere systems, smaller cells are more isotropic.
For very oblate ellipsoids with $\alpha = 0.60$ or $0.40$, the anisotropy in Fig.~\ref{fig:v-beta-jammed} appears to be rather independent of the cell volume.

\subsection{Poisson hyperplane tessellations}
\label{sec:PHP-STIT-local-beta}

\begin{figure}[t]
  \centering
  \includegraphics[width=0.8\textwidth]{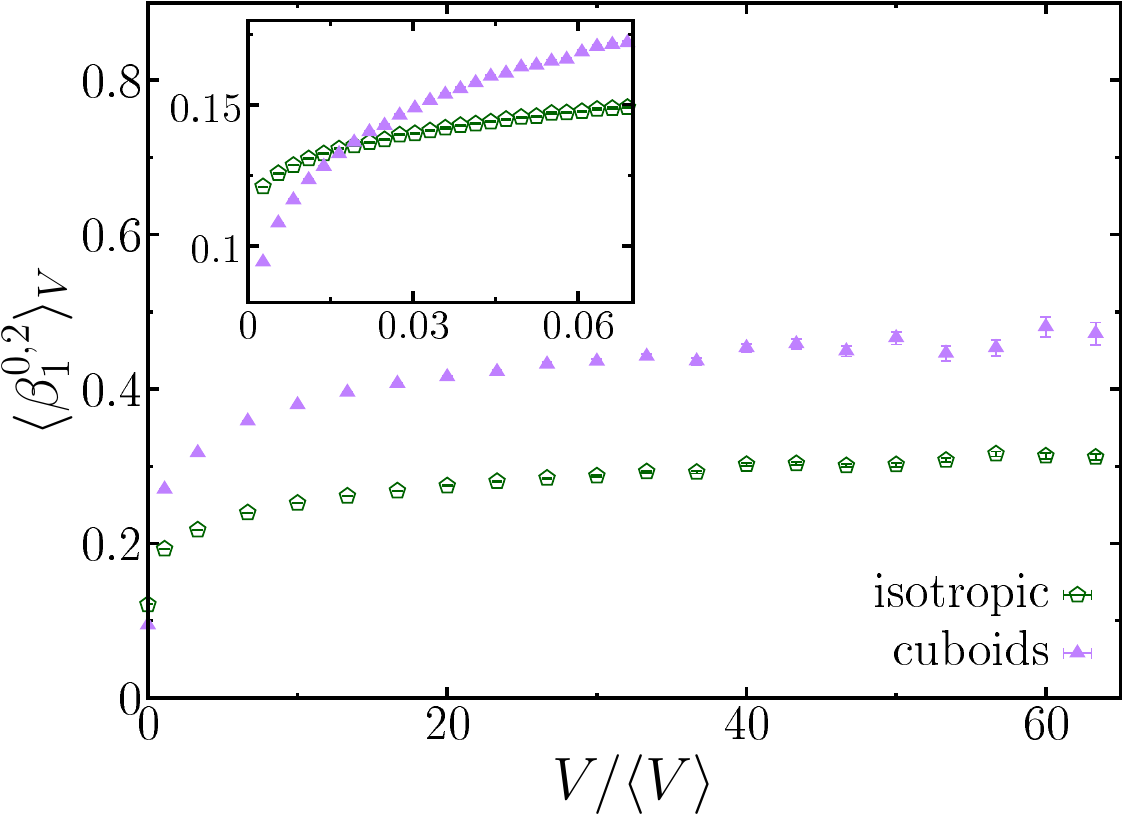}
  \caption{Anisotropy as a function of cell volume for PHP
    tessellations with either an isotropic
    orientation distribution of the hyperplanes or with three allowed
    directions, i.\,e., all cells are cuboids. For most volumes $V$, the cells in the
    isotropic tessellation are on average more anisotropic than the
    cuboid shaped cells. However, the inset shows that this changes for
    very small cells, for which the cells in the isotropic system are more
    isotropic than the cuboid-shaped cells.}
\label{fig:v-beta-PHP}
\end{figure}

Figure~\ref{fig:v-beta-PHP} displays the results for the PHP tessellations with either an isotropic orientation distribution of the hyperplanes or with three allowed directions, i.\,e., all cells are cuboids.

The main plot shows the anisotropy as a function of the cell volume for large cells, which get exceedingly unlikely with increasing cell volume, see Fig.~\ref{fig:epdf-PHP}.
These large cells are on average more isotropic than a typical cell in the tessellation, and the cuboid-shaped cells are more isotropic than the cells in the statistically isotropic tessellation (i.\,e., larger values of $\langle\beta_1^{0,2}\rangle_V$).
However, the inset shows that this order reverses for small cells, where the cells in the statistically isotropic system are more isotropic than the cuboid shaped cells.


\section{Conclusions}
\label{sec:conclusions}

Random or disordered tessellations appear in very different physical, chemical, or biological systems as well as in life sciences.
Their complex structure calls for advanced mathematical tools that can quantify their geometry.

The Minkowski functionals and tensors allow for a robust yet concise characterization of the shape of single cells.
They are powerful tools to narrow the choice of possible underlying stochastic processes.

Here, we have applied, in a theory-based simulation study, such an analysis to a variety of important and common tessellations, see Section~\ref{sec:pp_def}.
\begin{itemize}
  \item The ``map of anisotropy'' from Section~\ref{sec:map} analyzes the relationship between
the dimensionless ratio $\langle V\rangle^2/\langle A\rangle^3$ of average cell volumes to average cell areas
and the degree of cell elongation quantified by the eigenvalue ratio $\langle \beta_1^{0,2}\rangle$ of the interface Minkowski tensors $W_1^{0,2}$.
It provides an overview of the various tessellations considered here.
It can highlight relations between different point processes but also reveals some structural differences.
\item The probability density functions of single local characteristics in Section~\ref{sec:distributions} can also clearly distinguish two different types of tessellations such as the PHP and Voronoi tessellations.
However, the rescaled probability density functions for Voronoi cells from different stochastic processes can be qualitatively similar.
On the one hand, this agreement reveals interesting relations between the models.
On the other hand, it does not imply that the local structure, i.\,e., the shape distribution of single cells, is the same for these physically quite different point processes.
\item To detect the differences in the local structure, we combine different characteristics.
More precisely, we determine the mean anisotropy index as a function of the cell volume, see Section~\ref{sec:joint-characterization}.
Thus, we find a qualitatively different behavior, for example, for determinantal point processes and equilibrium hard spheres.
We also use this analysis for additional insights into the Poisson hyperplane or STIT tessellations, e.\,g., discussing the different anisotropy for small or large cells. 
The numerical tools which we apply here are efficient and can be easily used for a detailed structure analysis of any tessellation of interest.
\end{itemize}


We have thus demonstrated how the Minkowski functionals and tensors can serve as sensitive and robust local shape descriptors.
Given a simple object like a single cell and starting with simple and efficient shape indices like volume and surface area (following the rule of parsimony),
the straightforward generalization to Minkowski functionals and tensors allows for a comprehensive shape analysis.
Each additive, continuous, and motion invariant or covariant tensor is essentially a linear combination of Minkowski tensors~\cite{Hadwiger1951,Alesker:1999a,Alesker:1999b}.
Moreover, these geometrical shape descriptors are more robust than so-called ``topological measures,'' like the number of faces, vertices, or edges.
Such topological quantities are sensitive to noise in that a small change can strongly alter the topology of the cell.
For example, whether or not a small additional face is resolved can lead to faces with a very different number of vertices.

Prominent examples are also the so-called bond-orientational order parameters, which are standard tools in condensed matter physics to characterize particle arrangements~\cite{SteinhardtEtAl1983}.
They are based on the definition of a neighborhood, where different choices can even lead to qualitatively different behavior of the bond-orientational order parameters.
Moreover, because of the discrete nature of neighborhood, the bond-orientational order parameters can change discontinuously for infinitesimally small changes in the particle positions.
This can be avoided by a morphometric approach that assigns weights to the neighbors.
These weights lead to robust measures that are continuous in the particle coordinates and that are equivalent to the Minkowski tensors presented here.
%

\begin{acknowledgement}
  We thank Andy Kraynik, Markus Spanner, and Richard Schielein for their data of the
  monodisperse foam, the equilibrium hard-sphere liquids, and the crystalline sphere packings, respectively.
  We also thank Felix Ballani for his software simulating the typical cell of a Poisson hyperplane tessellation.
  We thank Markus Kiderlen for valuable discussions and suggestions.
  We also thank the German science foundation (DFG) for the grants HU1874/3-2, LA965/6-2, SCHR1148/3 and ME1361/11 awarded as part of the DFG-Forschergruppe ``Geometry and Physics of Spatial Random Systems''.
\end{acknowledgement}

\bibliographystyle{spmpsci}
\bibliography{cell-shapes}

\end{document}